\documentclass[english,aps,pra, twocolumn, floatfix, showpacs, hypref, superscriptaddress]{revtex4-1}
\usepackage{bm}
\usepackage{lipsum}
\usepackage{amsmath}
\usepackage{graphicx}
\usepackage{color}
\usepackage{hyperref}
\usepackage{epstopdf}
\makeatletter
\@ifundefined{textcolor}{}
{%
 \definecolor{BLACK}{gray}{0}
 \definecolor{WHITE}{gray}{1}
 \definecolor{RED}{rgb}{1,0,0}
 \definecolor{GREEN}{rgb}{0,1,0}
 \definecolor{BLUE}{rgb}{0,0,1}
 \definecolor{CYAN}{cmyk}{1,0,0,0}
 \definecolor{MAGENTA}{cmyk}{0,1,0,0}
 \definecolor{YELLOW}{cmyk}{0,0,1,0}
}

\begin{document}

\title{Chiral Bosonic Phases on the Haldane Honeycomb Lattice}

\begin{abstract}
Recent experiments in ultracold atoms and photonic analogs have reported the implementation of artificial gauge fields in lattice systems, facilitating the realization of topological phases. Motivated by such advances, we investigate the Haldane honeycomb lattice tight-binding model, for bosons with local interactions at the average filling of one boson per site. We analyze the ground state phase diagram and uncover three distinct phases: a uniform superfluid (\textit{SF}), a chiral superfluid (\textit{CSF}) and a plaquette Mott insulator with local current loops (\textit{PMI}). Nearest-neighbor and next-nearest neighbor currents distinguish \textit{CSF} from \textit{SF}, and the phase transition between them is first order. We apply bosonic dynamical mean field theory and exact diagonalization to obtain the phase diagram, complementing numerics with calculations of excitation spectra in strong and weak coupling perturbation theory. The characteristic density fluctuations, current correlation functions, and 
excitation spectra are measurable in ultracold atom experiments.

\end{abstract}
\author{Ivana Vasi\'c}
\affiliation{Institut f\"ur Theoretische Physik, Goethe-Universit\"at,
60438 Frankfurt/Main, Germany}
\author{Alexandru Petrescu}  
\affiliation{Department of Physics, Yale University, New Haven, CT 06520, USA}
\affiliation{Centre de Physique Theorique, Ecole Polytechnique, CNRS, 91128 Palaiseau Cedex, France}
\author{Karyn Le Hur} 
\affiliation{Centre de Physique Theorique, Ecole Polytechnique, CNRS, 91128 Palaiseau Cedex, France}
\author{Walter Hofstetter}
\affiliation{Institut f\"ur Theoretische Physik, Goethe-Universit\"at,
60438 Frankfurt/Main, Germany}
\maketitle

\section{Introduction}
\label{Sec:Intro}
Magnetic fields play a crucial role in condensed-matter physics, from the complete expulsion of magnetic fields in superconductors (Meissner effect) to the appearance of quantum Hall states. More generally, gauge fields play a central role in the description of macroscopic quantum phenomena. Lattice variants of the quantum Hall effect have attracted attention since the 1980s, beginning with the groundbreaking work by Hofstadter \cite{Hofstadter}, followed by a complete characterization of magnetic bands via topological quantum numbers \cite{TKNN}. In 1988, Haldane \cite{Haldane} introduced a fermionic tight-binding model on the honeycomb lattice that breaks time-reversal symmetry without net magnetic flux through the unit cell. The model exhibits non-trivial topological properties as a result of next nearest-neighbor tunneling processes. Time reversal symmetric extensions, 2D topological insulators \cite{KaneMele, BHZ} (for a review see \cite{Cayssol}) have been experimentally realized in HgTe quantum wells \cite{
Molenkamp}. Revived interest into these 
models   stems from on-going experiments in photonic lattices \cite{HaldaneRaghu,RaghuHaldane, MITphotons, Hafezi0, Hafezi, 
Rechtsman, Carusotto, Koch0, Petrescu0, Koch, Fang, BlochAmo, Lu}, metamaterials \cite{Khanikaev, Sommer} and optical lattices hosting ultracold atoms \cite{Spielman, Dalibard, Goldman, Struck, Aidelsburger, Miyake, ExpHaldane}. 

For ultracold atoms in optical lattices, artificial gauge fields producing complex hopping amplitudes have been realized  by shaking the optical lattice which results in a net Peierls phase \cite{Struck, ExpHaldane} or by laser--assisted tunneling \cite{Aidelsburger}. Two groups have recently reported the realization of a Hofstadter butterfly model \cite{Aidelsburger, Miyake}, and very recently the first experimental realization of the Haldane model has been achieved \cite{ExpHaldane}. Other ultracold atom experiments have also succeeded in realizing triangular flux lattices \cite{Struck}.
The current technologies allow to realize one-dimensional and two-dimensional  lattice systems which can be loaded with bosons or fermions. These recent developments constitute impressive steps towards simulating many-body physics, artificial gauge fields and spin-orbit couplings in optical lattices where interactions can be engineered and tuned in a precise manner.

\begin{figure}[!tbh]
\includegraphics[width=\linewidth]{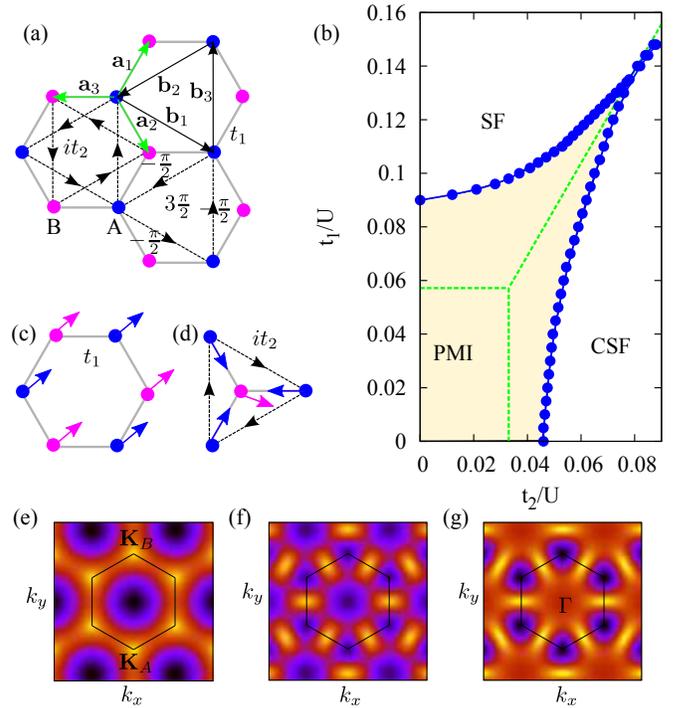}
\caption{(Color online) \textbf{a)} Lattice vectors on the honeycomb lattice and hopping integrals the Haldane model of Eq.~(\ref{eq:ham}). \textbf{b)} Phase diagram of the model (\ref{eq:hami}) at unit filling, containing \textit{plaquette Mott insulator} (\textit{PMI}), \textit{uniform superfluid} (\textit{SF}) and \textit{chiral superfluid} (\textit{CSF}) phases. Solid (dashed) lines represent DMFT (Gutzwiller mean--field) results.  \textbf{c)} Local condensate order parameter in the uniform superfluid; \textbf{d)} In CSF the condensate order parameters on sublattices $A$ and $B$ are determined up to a relative phase.   \textbf{e)}-\textbf{g)} The lowest band of Eq.~(\ref{eq:ham}) for $t_1=1$, $\phi=\pi/2$ and $t_2=0, 1/\sqrt{3}, 1$, from the left to the right. Band minima move from the center ${\bf \Gamma}$ to the corners $\mathbf{K}_A$, $\mathbf{K}_B$ of the first Brillouin zone (depicted as a solid line). At $t_1 = \sqrt{3} t_2$ there are three degenerate minima.}
\label{Fig:Fig1}
\end{figure}

In parallel, several theoretical works have focused on  interaction--induced transitions from a topological into a Mott insulator (MI) in fermionic systems \cite{RaghuTMI,Varney,PesinBalents, Rachel,  WuRachel, Hohenadler, Hohenadler2, Ruegg, Cocks, OrthCocks, Liu}. Properties of lattice bosons exposed to artificial gauge fields have been addressed as well in different regimes. Lattices with staggered flux give rise to finite momentum Bose--Einstein condensates \cite{Lim1, Lim2, Moller}. A related phenomenon has been identified in the presence of uniform flux and the excitation spectrum in the weakly interacting regime of this unusual superfluid phase has been calculated \cite{Powell1, Powell2}. With stronger interactions, a superfluid to MI transition is expected to occur \cite{OriginalSFMI, Greiner, Lim1, Lim2, Niemeyer, Oktel, Mueller, Sengupta2, Yao}, and even more interestingly, the interplay of strong interactions and uniform lattice flux should lead to fractional Hall states  \cite{Regnault} and 
topological transitions \
\cite{Huber}. Another proposed setup 
for reaching the 
quantum Hall regime are optical flux lattices \cite{Cooper}. In low dimensional lattices with 
staggered flux a new intermediate phase has been predicted - a chiral Mott state \cite{Paramekanti1, Paramekanti2, Zaletel} that exhibits  broken time--reversal symmetry without breaking  $U(1)$ symmetry. Bond-chirality and plaquette order have been shown to emerge in a system of two-dimensional hard-core bosons with frustrated ring exchange \cite{Huerga}. Topological 
transport in 
bosonic Mott states in the presence of spin orbit--coupling has been studied in \cite{WongDuine1,WongDuine2}. Emergence of a chiral current and Meissner effect in bosonic ladders have been demonstrated experimentally \cite{ Atala} and theoretically analyzed \cite{OrignacGiamarchi, Petrescu, Paredes, Georges, Wei}. While topological bosonic Mott insulators have been theoretically predicted in one dimension \cite{Duan, Grusdt, Santos}, an important open question in the field is the existence of  bosonic topological Mott states in 2 spatial dimensions \cite{Wen}. Multicomponent interacting bosonic systems  exhibit spontaneous spin Hall effect \cite{CSSF}, exotic magnetic order \cite{Piraud} and integer Hall effect \cite{Regnault2}. A low density ground state of spinor bosonic gases with isotropic Rashba spin--orbit coupling is proven to be a composite fermion state \cite{Sedrakyan}.

Recently several approaches for the realization of the Haldane model for integer quantum Hall effect without Landau levels \cite{Haldane} in an ultracold atom system were theoretically proposed \cite{Wang, Stanescu, Goldman2, Anisimovas}. In relation to this, the intricacies of the direct Peierls substitution for the Haldane model were addressed \cite{Ibanez}. The very recent experiment \cite{ExpHaldane} demonstrates that time-periodic driving of a honeycomb optical lattice creates the prerequisite complex next-nearest neighbor hopping. Topological transitions arising in the non-interacting Haldane model for fermions have been directly probed in this experimental setup.

Motivated by experimental possibilities and open theoretical questions on the emergence of new bosonic phases in the presence of gauge fields, we study the Haldane model for bosons at unit filling with a local repulsive Hubbard interaction. The Hamiltonian comprises three terms: a real nearest neighbor hopping $t_1$, an imaginary next-nearest neighbor hopping $i t_2$ and the local repulsive interaction $U$. Each of the three terms favors one of the phases depicted in the diagram of Fig.~\ref{Fig:Fig1}b, which is our main result. The $t_1$-dominated phase is a \textit{uniform superfluid} (\textit{SF}) with long range phase correlations, whereas the $t_2$-dominated phase is a \textit{chiral superfluid} (\textit{CSF}) that exhibits phase modulation due to bosons condensing at nonzero momentum. The interactions dominated phase is a \textit{plaquette Mott insulator} (\textit{PMI}) characterized by local plaquette currents. We find that the excitation bands in both superfluid and Mott insulator phases are 
reminiscent of the single-particle spectrum of the Haldane model. For all 
considered phases we calculate experimentally accessible features such as density fluctuations, plaquette currents and excitation spectra.

The paper is organized as follows: we introduce the Haldane--Hubbard model in Sec.~\ref{Sec:NImodel}. We discuss the weakly interacting limit and the distinction between the uniform superfluid and the chiral superfluid in Sec.~\ref{sec:SecWIB}, followed by a discussion of possible ground states and current expectation values in Sec.~\ref{Subsec:CrtWeak}. To access the regime of stronger interactions, we start with a simple mean--field theory approach that leads to a phase diagram which captures all the important qualitative features in Sec.~\ref{Subsec:Gutzwiller}. To address effects beyond the mean field approximation, in Sec.~\ref{Subsec:DMFT} we use bosonic dynamical mean--field theory (DMFT) \cite{bdmft1, bdmft2, Hubener, Snoek&Hofstetter, Anders} that gives information on the thermodynamic limit.  We then compare such results to those extracted from the exact ground state found with the Lanczos algorithm for a finite system in Sec.~\ref{Subsec:Lanczos}. Within all these approaches, we compute the 
values of 
condensate order parameters, density fluctuations and 
plaquette currents in the ground state. We then turn to the excitation spectra of the superfluid (Sec.~\ref{Subsec:EXCSF}) and of the Mott insulator (Sec.~\ref{Subsec:EXCMI}) and characterize their main features. We conclude in Sec.~\ref{Sec:Conclusions} with a discussion of our results and indicate possible future research directions.

\section{Model}
\label{Sec:NImodel}
The Haldane Hamiltonian on the honeycomb lattice is given by \cite{Haldane}
\begin{equation}
\mathcal{H}_{\mathrm{H}} = - t_1 \sum_{\langle i,j \rangle}\hat b_i^{\dagger} \hat b_j- t_2 \sum_{\langle\langle i,j\rangle\rangle}e^{i \phi_{ij}} \hat b_i^{\dagger} \hat b_j,
\label{eq:ham}
\end{equation}
where $\hat{b}_i$ is the annihilation operator at site $i$. The term proportional to $t_1$ describes the graphene lattice with non--zero hopping elements only between nearest neighbors along vectors $\bf{a_1}$, $\bf{a_2}$ and $\bf{a_3}$ shown in Fig.~\ref{Fig:Fig1}a). The $t_2$ term was originally introduced by Haldane. It includes a complex phase $\phi_{ij}$ for the tunneling between two next--nearest neighbors (sites belonging to the same sublattice) along $\bf{b_1}=\bf{a_2}-\bf{a_3}$, $\bf{b_2}=\bf{a_3}-\bf{a_1}$, and $\bf{b_3} = \bf{a_1}-\bf{a_2}$. The absolute value of $\phi_{ij}$ is constant throughout the lattice and its sign is shown in Fig.~\ref{Fig:Fig1}a). As can be seen from the same figure, the net flux per unit cell is zero. In the following, we consider the case of $|\phi_{ij}|=\pi/2$ and use the notation $t_{ij}$ to shorten Eq.~(\ref{eq:ham}) to $ \mathcal{H}_{\mathrm{H}} = -\sum_{i,j} t_{ij} \hat b_i^{\dagger} \hat b_j$. We will add a label to make sublattice dependence 
explicit when necessary, for example $\hat b_i \rightarrow \hat b_{A i}$. 

The single particle Hamiltonian~(\ref{eq:ham}) can be described by the Chern numbers of its Bloch bands, a property which we briefly review here. In momentum space, (\ref{eq:ham}) is rewritten as
$
 \mathcal{H}_{\mathrm{H}} = \int_\text{BZ} d{\bf k}\, \psi({\bf k})^{\dagger}\mathcal{H}_{\mathrm{H}}({\bf k})\psi({\bf k}), 
$ with
\begin{equation}
\mathcal{H}_{\mathrm{H}}({\bf k}) = -{\bf d}\,(\bf k) \cdot { \hat {\bf \sigma}}.
\label{eq:hamk}
\end{equation}
The momenta $\mathbf{k}$ belong to the first Brillouin zone, which is spanned by the vectors $\mathbf{g}_1 = \left(2\pi / a, - 2\pi/(a\sqrt{3})\right)$ and $\mathbf{g}_2 = \left(0, 4 \pi/(a\sqrt{3}) \right)$ in reciprocal space. We have introduced the field $\psi({\bf k})=\left(b_A({\bf k}), b_B({\bf k})\right)^T$ of Fourier transforms of the annihilation operators on sublattices $A$ and $B$. We wrote $\mathcal{H}_\text{H}$ in the basis of Pauli matrices $\hat{\mathbf{\sigma}} = (\sigma_x, \sigma_y, \sigma_z)$ in terms of  
\begin{equation}
{\bf d}({\bf k})=\left( 
t_1 \sum_i\cos{\bf k}\, {\bf a_i},
t_1 \sum_i\sin{\bf k}\, {\bf a_i}, 
-2 t_2 \sum_i\sin{\bf k}\, {\bf b_i}
\right).\label{eq:d}
\end{equation} 
The non-trivial topology of the Bloch bands translates to a nonzero winding number of the map $ \hat{\mathbf{d}} = \mathbf{d} / |\mathbf{d}|$  from the torus (the first Brillouin zone) to the unit sphere. Denoting by $\partial_{i}$ the partial derivatives with respect to the two components of momentum $k_i$, $i = 1,2$, the winding number is \cite{Haldane}
\begin{equation}
\mathcal{C}_- =  \frac{1}{4\pi} \int_\text{BZ} d\mathbf{k}\, \hat{\mathbf{d}} \cdot \left( \partial_1 \hat{\mathbf{d}} \times \partial_2 \hat{\mathbf{d}} \right).
\label{eq:ti}
\end{equation}
Eq.~(\ref{eq:ti}) represents the Chern number of the lower Bloch band, and it takes the value $\mathcal{C_-} = 1$ at finite values of $t_2$ and $t_1$ when $|\phi_{ij}| = \frac{\pi}{2}$. For $t_1=0$ the two bands touch along the certain cuts of the Brillouin zone, and the spectrum is fully gapped as soon as $t_1 > 0$. The formula for the upper band is obtained by replacing $\hat{\mathbf{d}}$ by $-\hat{\mathbf{d}}$, and leads to $\mathcal{C}_+ = -1$.

We focus here on the bosonic Haldane--Hubbard Hamiltonian
\begin{equation}
\mathcal{H}=\mathcal{H}_{\mathrm{H}} + \frac{U}{2}\sum_i \hat n_i \left(\hat n_i -1\right) -\mu\sum_i \hat n_i,
\label{eq:hami}
\end{equation}
where $U$ is a local (on--site) interaction and $\mu$ is the chemical potential. Throughout this work, we will consider the zero temperature limit $T=0$. In the following sections, we characterize different phases by the value of the condensate order parameter
\begin{equation}
 \psi_i = \langle \hat b_i\rangle,
\end{equation}
local density fluctuations
\begin{equation}
 \Delta n_i = \langle \hat n_i^2\rangle -\langle \hat n_i \rangle^2,
\end{equation}
and emerging patterns of lattice currents. The expectation value of the current operator for the bond $j\rightarrow i$ on the lattice is given by
\begin{equation}
  J_{ij}= -i (t_{ji} \langle \hat b_j^\dagger \hat b_i \rangle- t_{ij} \langle \hat b_i^\dagger \hat b_j \rangle) =  -2 \mathrm{Im}\left(t_{ij} \langle \hat b_i^\dagger \hat b_j \rangle\right),
\label{eq:jij}
\end{equation}
as can be derived from the lattice continuity equation.

 Implications of non--trivial Chern numbers (\ref{eq:ti}) are most often discussed in the context of fermionic systems.
 In the seminal paper \cite{TKNN}, the Hall conductance of non--interacting lattice fermions in the strong magnetic field has been expressed in terms of Chern numbers of occupied Bloch bands. 
 The current cold--atom realization of the Haldane model \cite{ExpHaldane} is based on the idea of Floquet topological insulators \cite{Lindner, *Kitagawa, *Cayssol2}. The definition and the meaning of topological invariants in these periodically driven systems 
 have been in the focus of several studies \cite{Rudner, Carpentier, Mitra}.
 Recently, the first photonic analogs of topologically non--trivial systems have been realized \cite{Lu}: both in the classical regime \cite{MITphotons, Rechtsman} and in the quantum regime where arrays of coupled photonic cavities have been used \cite{ Hafezi}. 
 These photonic experiments have probed the emerging edge states, as a clear indication of non--trivial topology. More recently, it  has been theoretically proposed to directly measure topological invariants \cite{Petrescu0, Ozawa, HafeziPRL2014} in these systems. Topological transitions have been directly probed in quantum circuits of interacting  superconducting qubits \cite{Roushan}.

Photonic systems \cite{Hafezi, Carusotto} typically work in the dissipative--driven regime.  In the equilibrium situation that we consider througout the paper, 
when a topological band is filled with weakly interacting bosons, the ground--state can be a topologically trivial Bose--Einstein condensate \cite{LeBlanc03072012, Powell1, Powell2, Price}.  
Properties of the condensate are set by the features of band minima and are not affected by a non--trivial band topology expressed by equation (\ref{eq:ti}). 
However, band topology does affect transport properties of bosons as well as properties of excitations \cite{LeBlanc03072012, Powell1, Powell2, Price, WongDuine1, WongDuine2}. Transport of lattice bosons has been used to probe a finite Berry curvature \cite{Duca} and the Chern number of a topological band \cite{Aidelsburger2014}. In contrast to weakly--interacting bosons, strongly interacting (hard--core) bosons in topological flat bands at certain filling fractions are known to exhibit topologically nontrivial ground states \cite{Sheng}. In the next sections, we study the ground state and excitations of the model (\ref{eq:hami}).

\section{Study of the ground state}
\subsection{Weakly interacting bosons}
\label{sec:SecWIB}

In the weakly interacting limit, we expect bosons to condense. From the non--interacting model of Eq.~(\ref{eq:ham}), we can easily
infer two  limits that give rise to two types of superfluids: for $t_2=0$ we obtain a honeycomb lattice and all bosons condense at the center of the Brillouin zone ${\bf \Gamma}$ at zero momentum. On the other hand, for $t_1=0$, the model turns into two decoupled triangular lattices, and we expect separate condensation of bosons on two sublattices.

To study the possible condensates at finite values of $t_1$ and $t_2$, let us focus on the non--interacting Hamiltonian. The energy dispersion of the lowest band $\epsilon_ -({\bf k})$ exhibits either a single minimum at ${\bf k}=\mathbf{\Gamma}$ for $t_1>\sqrt{3} t_2$ or degenerate minima at the two inequivalent corners of the Brillouin zone ${\bf K}_{A} $ and ${\bf K}_{B}$ for $t_1 < \sqrt{3} t_2$ [see Fig.~\ref{Fig:Fig1}e)-g)]. The momenta at the corners of the Brillouin zone satisfy 
\begin{equation}
  e^{i {\bf K}_{A} \cdot {\bf b}_i}=e^{i \frac{2\pi}{3}}, \quad e^{i {\bf K}_{B} \cdot {\bf b}_i} =e^{-i \frac{2\pi}{3}},
 \label{eq:kpoints}
\end{equation}
for all $i=1,2,3$. At these high symmetry points, the Hamiltonian takes the following forms: 
\begin{equation}
\mathcal{H}(\mathbf{\Gamma}) = -3 t_1 \sigma_x
\end{equation}
and 
\begin{equation}
\mathcal{H}(\mathbf{K}_{A,B}) = \pm 3 \sqrt{3}t_2 \sigma_z.
\label{eq:corner}
\end{equation}

The \textit{SF} phase forms when $t_1>\sqrt{3} t_2$. The condensate order parameter is $\langle \hat b_{i} \rangle =\sqrt{n}$, where $n = N / N_\text{sites}$ is the filling. The ground state energy obtained from the Gross--Pitaevskii (GP) energy functional \cite{GPbook}
\begin{equation}
E_0 = -\sum_{i,j} t_{ij} \psi_i^* \psi_j + \frac{1}{2} U \sum_i |\psi_i|^4
\label{eq:gpenergy}
\end{equation}
is $\frac{E_0}{N_\text{sites}} = \left(-3 t_1 n + \frac{1}{2}U n^2\right)$ and we find that the next--nearest neighbor hopping is effectively cancelled. Using Eq.~(\ref{eq:jij}), the next--nearest neighbor bond current is 
\begin{equation}
J^\text{\textit{SF}}_{AA} = -2\,n\,t_2\, \mathrm{Im}\exp (-i\pi/2) = 2 n t_2.\label{eq:current1}
\end{equation}

The \textit{CSF} phase forms in the opposite case $t_1< \sqrt{3} t_2$. Non--interacting bosons can condense in a state that is an arbitrary linear combination of single particle ground states at ${\bf K}_{A}$ and ${\bf K}_{B}$, leading to large degeneracy. However, even weak repulsive interactions prefer a uniform density distribution on the two sublattices. To infer the low energy description,  we assume that only the minima of the lowest band are occupied and approximate operators according to Eq.~(\ref{eq:corner}) by
\begin{eqnarray}
 \hat b_{A, i} &\approx& \frac{1}{\sqrt{N_\text{sites}/2}} e^{-i {\bf K}_A {\bf r}_i} \hat b_{A}({\bf K}_A), \nonumber\\ \hat b_{B, i} &\approx& \frac{1}{\sqrt{N_\text{sites}/2}} e^{-i {\bf K}_B {\bf r}_i} \hat b_{B}({\bf K}_B).
\end{eqnarray}
The Hamiltonian (\ref{eq:hami}) then turns into
\begin{eqnarray}
\mathcal{H}&\approx&-3\sqrt{3} t_2\left(\hat b^{\dagger}_{A}({\bf K}_A) \hat b_{A}({\bf K}_A)+\hat b^{\dagger}_{B}({\bf K}_B) \hat b_{B}({\bf K}_B)\right)\nonumber\\&+&\frac{U}{2} b^{\dagger}_{A}({\bf K}_A) \hat b_{A}({\bf K}_A)\left(\frac{2}{N_{\text{sites}}}b^{\dagger}_{A}({\bf K}_A) \hat b_{A}({\bf K}_A)-1\right)\nonumber\\ &+&\frac{U}{2}b^{\dagger}_{B}({\bf K}_B) \hat b_{B}({\bf K}_B)\left(\frac{2}{N_{\text{sites}}}b^{\dagger}_{B}({\bf K}_B) \hat b_{B}({\bf K}_B)-1\right)\nonumber\\&-&\mu \left(\hat b^{\dagger}_{A}({\bf K}_A) \hat b_{A}({\bf K}_A)+\hat b^{\dagger}_{B}({\bf K}_B)\hat b_{B}({\bf K}_B)\right),
\label{eq:dw}
\end{eqnarray}
\textit{i.e.}~it describes two decoupled sublattices since the nearest--neighbor tunneling term vanishes
\begin{equation}
\sum_{\langle i,j\rangle}\hat b^{\dagger}_i\hat b_j \propto\sum_i \hat b^{\dagger}_i \sum_{j=1}^{3} e^{-i {\bf K}_B \cdot ({\bf r}_i+{\bf a}_j)}\hat b_{B}({\bf K}_B)=0.
\label{eq:decoupling}
\end{equation}
Thus we conclude that the ground state consists of two decoupled superfluids.  The same observation follows directly from Eq.~(\ref{eq:gpenergy}),  i.~e.~at the mean--field level the ground--state consists of two separate condensates occupying the two sublattices. 

The mean-field ground state energy (\ref{eq:gpenergy}) is $\frac{E_0}{N_{\text{sites}}} = \left(-3 \sqrt{3} t_2 n + \frac{1}{2}U n^2\right)$ and the corresponding momentum distributions are $\rho_A({{\bf k}})\approx \frac{N}{2}\,\delta_{{\bf k},{\bf K}_{A}}$, $\rho_B({{\bf k}})\approx \frac{N}{2}\,\delta_{{\bf k},{\bf K}_{B}}$. For the operators $\hat b_{A,i}$ on the same sublattice we find from Eq.~(\ref{eq:kpoints}) 
 \begin{eqnarray}
 \langle \hat b^{\dagger}_{A,i} \hat b_{A,j}\rangle&=&\psi^*_{A,i} \psi_{A,j}=\frac{2}{N_{\text{sites}}}\sum_{\bf k} e^{i {\bf k}({\bf r_i}-{\bf r_j})} \langle \hat b^{\dagger}_{A}({\bf k}) \hat b_{A}( {\bf k})\rangle\nonumber\\&=&n\exp\left( i \frac{2\pi}{3}m\right),
 \end{eqnarray}
  where $m$ is an arbitrary integer.
 The condensate at nonzero momentum exhibits nonuniform phase differences between next-nearest neighbors [see Fig.~\ref{Fig:Fig1}d)]. Phase ordering directly affects the next--nearest neighbor current expectation value  
\begin{eqnarray}
J^\textit{CSF}_{AA} &=& -2 ~\text{Im} \left( t_2 e^{-i\pi/2} \left\langle \hat{b}^\dagger_{Ai} \hat{b}_{Aj}    \right\rangle \right) \nonumber \\
&=& - 2 t_2 n \sin\left[ -\pi/2 + \mathbf{K}_A \cdot (\mathbf{r}_i - \mathbf{r}_j) \right]=- n t_2.\label{eq:current2}
\end{eqnarray} 
 The aforementioned ``decoupling of sublattices", Eq.~(\ref{eq:decoupling}), is depicted in Fig.~\ref{Fig:Fig1}d as an arbitrary phase difference between order parameters on two sublattices.

At the critical hopping strength $t_1=\sqrt{3}t_2$, there are three degenerate minima present in the dispersion relation, and in order to deduce the proper ground state at the mean--field level all three of them should be taken into account. However, the analysis of the mean-field energy functional indicates that condensation either at ${\bf \Gamma}$ or at both ${\bf K}_A$ and ${\bf K}_B$ is preferred, and we do not find a density modulated phase in this case \cite{Powell1, Powell2}.

The fact that at the phase boundary between the two superfluids the current changes abruptly, Eqs.~(\ref{eq:current1}) and (\ref{eq:current2}), hints towards a first order phase transition between uniform and chiral superfluids in the weakly interacting limit.

\subsection{Josephson effect between sublattices}
\label{Subsec:CrtWeak}
In Subsec.~\ref{sec:SecWIB} we distinguished \textit{SF} and \textit{CSF} phases through their patterns of the next--nearest neighbor current $J_{AA}$ (equivalently $J_{BB}$), expressed in Eqs.~(\ref{eq:current1}) and~(\ref{eq:current2}), respectively.  We now argue that the two superfluid phases have different expectation values of the nearest neighbor current $J_{AB}$, which is a signature of Josephson-type phase coherence between sublattices $A$ and $B$.    

In the \textit{SF} phase, the phases of bosons on the $A$ and $B$ sublattices are pinned and therefore the nearest neighbor current vanishes
\begin{equation}
J^\textit{SF}_{AB} = 0.
\end{equation} 
This follows from the fact that the operator corresponding to the boson at the minimum of the lower band is $\hat{b}_{-}( \mathbf{k} =  \mathbf{\Gamma} ) = \frac{1}{\sqrt{2}}[ \hat{b}_{A}(\mathbf{\Gamma}) + \hat{b}_{B}(\mathbf{\Gamma}) ]$. The ground state energy is invariant to a $U(1)$ rotation of the pinned phases on $A$ and $B$ sublattices, which corresponds to the existence of one Goldstone mode.

In the \textit{CSF} phase, the boson annihilation operator corresponding to the band minimum obeys $\hat{b}_{-}(\mathbf{K}_a) = \hat{b}_{a}(\mathbf{K}_a)$ for $a = A$ or $B$. The twofold degeneracy of the band minimum leads us to the problem of coherence of a Bose-Einstein condensate in a double-well potential \cite{Leggett, CSSF}. In the following, we prove that the presence of defects, or open boundary conditions, produces a condensate in which $A$ and $B$ sublattices are phase coherent. Secondly, we show that if discrete lattice symmetries are preserved, the ground state for weak $U>0$ consists of decoupled condensates on sublattices $A$ and $B$.

We form first a condensate wavefunction from \textit{coherent superpositions} of the degenerate minima
\begin{equation}
|\Phi_\textit{CSF}' ( \phi ) \rangle = \frac{1}{\sqrt{N!}} \left[\frac{1}{\sqrt{2}}\hat{b}^\dagger_{A}(\mathbf{K}_A) + \frac{e^{i\phi}}{\sqrt{2}} \hat{b}^\dagger_{B}(\mathbf{K}_B) \right]^N | 0 \rangle.
\label{eq:gs1}
\end{equation}
The nearest--neighbor current $J_{AB}$ on the unit cell at coordinate $\mathbf{r}_i$ has a well defined value $\propto \sin\left[ \phi - (\mathbf{K}_A - \mathbf{K}_B)\cdot \mathbf{r}_i \right]$. Note that the ground state $|\Phi_\textit{CSF}'\rangle$ spontaneously breaks lattice translation and lattice inversion symmetries.

We form a second wavefunction for the chiral superfluid by uniformly superimposing all $|\Phi_\textit{CSF}'(\beta)\rangle$ for $\beta$ between $0$ and $2\pi$. This new wavefunction corresponds to \textit{decoupled condensates}, and is both lattice translation and inversion symmetric:
\begin{equation}
| \Phi_\textit{CSF}'' \rangle = \frac{1}{(N/2)!} \left[\hat{b}^\dagger_{A}(\mathbf{K}_A) \right]^{N/2} \left[ \hat{b}^\dagger_{B}(\mathbf{K}_B) \right]^{N/2} | 0 \rangle.
\label{eq:gs2}
\end{equation}
Note that now the nearest neighbor current $J^\textit{CSF}_{AB}$ vanishes. This is due to the fact that the phases of the two sublattices can be rotated independently without changing the energy, which corresponds to the existence of two Goldstone modes.  

\begin{figure}[!tbh]                 
\includegraphics[width=\linewidth]{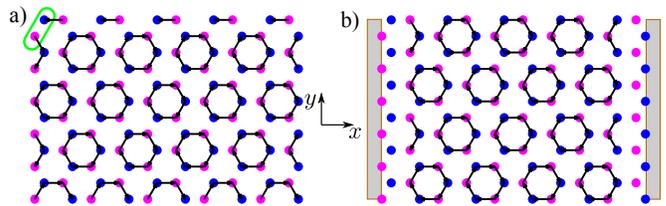}
\caption{(Color online) Aligning the phases of order parameters on the two sublattices: $U=1$, $t_1=10, t_2=10$, and average filling $n= 1$. In a) the top left link hosts a defect $t_1=4 t_2$. In b) we impose open boundary conditions in $x$ direction.
Arrows represent local (plaquette) currents with amplitude $|J_{\mathrm{AB}}|\approx |2 t_1 n \sin \frac{2\pi}{3}|= \sqrt{3} n t_1 $. Weaker currents that should vanish in the bulk, far away from the defect are not plotted. Results are obtained by minimizing the energy functional (\ref{eq:gpenergy}) with respect to $\psi_i$ for the 120 sites shown in the plot. In both plots there are density modulations - for example in a) there are more particles sitting on the sites linked by $4 t_1$.}
\label{Fig:Figgporder}
\end{figure}

 First we discuss a finite size system with periodic boundary conditions.
We find that the energy per site of $|\Phi_\textit{CSF}''\rangle$ is $\frac{E_0''}{N_\text{sites}} = - 3 \sqrt{3} t_2 n + (Un/2) \left( n + 1 - 2 / N_\text{sites} \right)$. This is lower by $n U /(2 N_\text{sites})$ than the energy of Eq.~(\ref{eq:gs1}). Thus, if the system is finite, if the interactions are weakly repulsive and if all discrete lattice symmetries are preserved, then the variational ground state is $|\Phi_\textit{CSF}''\rangle$. This is confirmed numerically using Lanczos methods for small translation and inversion symmetric clusters in Subsec.~\ref{Subsec:Lanczos}.  However, the fact that the states (\ref{eq:gs1}) and (\ref{eq:gs2}) become degenerate in the thermodynamic limit opens up a possibility of a ground state that breaks lattice symmetries \cite{Moller}. We investigate this issuse further in section \ref{Subsec:Lanczos} and \ref{Subsec:EXCSF}.

The ground state is significantly different if a defect is introduced in a finite lattice  with periodic boundaries or in a finite lattice with open boundaries. In these cases the double--well structure (\ref{eq:dw}) does not apply anymore. For example, at the boundary of the lattice, $A$ and $B$ sublattice phases can be pinned since the number of $B$ neighbors for any $A$ site is 2 instead of 3. Once the phases are pinned at the boundary, the $A$--$B$ sublattice phase coherence proliferates into the bulk. Another possibility to establish phase coherence between $A$ and $B$ sublattices is to create a strong nearest neighbor bond at a given unit cell (possibly imprinting a phase difference). As a consequence of long-range correlations between sites on the same sublattice, $J_{AB}$  at any other bond acquires a definite value. 

We conclude that 
\begin{equation}
J_{AB}^\textit{CSF} = 0
\end{equation}
in a finite system obeying lattice translation and inversion symmetries (\textit{i.e.} a lattice on a torus with finitely many sites), whereas
\begin{equation}
J_{AB}^\textit{CSF} \neq 0
\end{equation}
in the presence of defects or if the system has open boundaries, \textit{i.e.} in realistic experimental conditions. Our qualitative remarks about the role of defects are substantiated with numerical results obtained by minimizing the GP energy functional (\ref{eq:gpenergy}) for finite lattices, presented in Fig.~\ref{Fig:Figgporder}.

\subsection{Mott Insulator with Plaquette Currents}
\label{Subsec:Gutzwiller}
To address stronger interactions, we start with
  a mean--field decoupling  of the tunneling term as $\hat b_i^{\dagger} \hat b_j \approx \langle\hat b_i^{\dagger} \rangle\hat b_j+\hat b_i^{\dagger} \langle\hat b_j\rangle-\langle\hat b_i^{\dagger}\rangle \langle\hat b_j\rangle$ 
that is fully equivalent to applying a variational Gutzwiller ansatz $ |\psi_{\mathrm{GW}}\rangle=\prod_{\otimes i} \sum_{n} c_{i, n}|n\rangle$ \cite{Bissbort, KrutitskyNavez}. 
We obtain a mean--field Hamiltonian 
\begin{equation}
 \mathcal{H}_{\mathrm {mf}} = \sum_i  \mathcal{H}_{\mathrm {mf}}^{i} + \mathrm{const.},
 \label{eq:mfh}
\end{equation}
given by a 
sum of local terms
\begin{equation}
 \mathcal{H}_{\mathrm {mf}}^{i} = -\Psi_i^{\mathrm{mf}}\hat b^{\dagger}_i - \left(\Psi_i^{\mathrm{mf}}\right)^*\hat b_i+\frac{U}{2} \hat n_i \left(\hat n_i -1\right)-\mu\, \hat n_i,
 \label{eq:localmfh}
\end{equation}
where each lattice site is coupled to its neighbors
only by a sum of condensate order parameters $\Psi_i^{\text{mf}} = \sum_j t_{ij} \psi_j $. The ground state is found by  minimizing the expectation value of the Hamiltonian~(\ref{eq:mfh}) with respect to the coefficients $c_{i,n}$. The approximation becomes an exact description both in the weakly interacting limit and in the atomic limit. A resulting mean--field phase diagram at unit filling is shown in Fig.~\ref{Fig:Fig1}b by dashed lines. It consists of the two superfluid phases discussed in the previous subsection and in addition it contains a Mott phase. The rectangular shape of the Mott domain implies that the ground state energy of the uniform superfluid is unaffected by $t_2$ and vice versa for the chiral superfluid.

However, the Mott state that we obtain within the mean--field approximation, given by Eq.~(\ref{eq:mfh}) and (\ref{eq:localmfh}), is a simple product state $\prod_{\otimes i} |n\rangle$ and is therefore featurless. To explore its properties in more detail we need a better approach.
The Random Phase Approximation (RPA) \cite{Stoof, Sengupta, Freericks1} is analytically tractable. It is a standard approach that shares some similarities with DMFT and we briefly outline it here.

In both approaches, we consider the single--particle  Green's functions 
\begin{equation}
G_{ij}(\tau_1-\tau_2)=-\langle  \mathcal{T} \hat b_i(\tau_1) \hat b_j^{\dagger}(\tau_2)\rangle
\end{equation}
expressed in terms of Matsubara frequencies 
$\omega_n=2\pi n/\beta$, where $\beta $ is the inverse temperature ($\beta\rightarrow\infty$) and $G_{ij}(i\omega_n) = \int\, d{\tau} \exp(i \omega_n \tau) G_{ij}(\tau)$. The main approximation of the two methods is that the self--energy is local. This is an exact property in the limit of infinite lattice coordination number.

In RPA local self--energies are determined from the local Hamiltonian (\ref{eq:localmfh}) \cite{Stoof, Sengupta, Freericks1}. Local Green's functions corresponding to the  Hamiltonian (\ref{eq:localmfh}) at $T=0$ when the ground state is $\prod_{\otimes i} |n\rangle$ are given by
\begin{equation}
 \mathcal{G}_{ii}^{\mathrm{RPA}}(i\omega_n) = -\frac{n}{-(n-1) U +\mu + i \omega_n}+\frac{n+1}{-n U+\mu+i\omega_n}.
 \label{eq:gii}
\end{equation}
The corresponding self--energies are then calculated using the local Dyson equation
\begin{equation}
 \Sigma^{\mathrm{RPA}}_{i}(i\omega_n) = i\omega_n +\mu -  \left(\mathcal{G}_{ii}^{\mathrm{RPA}}(i \omega_n)\right)^{-1}.
 \label{eq:rpaselfenergy}
\end{equation}
As already mentioned, the last result approximates the self--energy of the full lattice problem and is used in the lattice Dyson equation, written either in real 
\begin{equation}
\left[ G^{\mathrm{RPA}}\right]^{-1}_{ij}(i\omega_n) =  \left(i\omega_n +\mu\right) \delta_{ij} + t_{ij} -  \delta_{ij}\Sigma_{i}^{\mathrm{RPA}}(i\omega_n),
 \label{eq:realspacedysonsc}
\end{equation}
or  $k$-space
\begin{equation}
\left[ G^{\mathrm{RPA}}\right]^{-1}(i\omega_n, {\bf k}) = \left(i\omega_n +\mu- \Sigma_{ii}^{\mathrm{RPA}}(i\omega_n)\right) \mathcal{I} -\mathcal{H}_{\mathrm H}( {\bf k}).
\label{eq:kspacedysonsc}
\end{equation}

Starting from the Green's functions, Eq.~(\ref{eq:kspacedysonsc}), we first derive the excitation spectrum of the Mott state.
By going into the basis in which $\mathcal{H}_{\mathrm H}( {\bf k}) $ is diagonal
and by applying analytical continuation $i\omega_n\rightarrow\omega+i\delta$, we can read off poles of the $\bf{k}$-dependent Green's function. The excitation spectrum of the Mott insulator is given by the particle--hole excitations
\cite{Stoof, Sengupta, KrutitskyNavez}
\begin{eqnarray}
  \omega_{\pm, \alpha} ({\bf k})&=& \frac{U}{2}\left((2 n-1)-2\frac{\mu}{U}+\frac{\epsilon_{\alpha}({\bf k})}{U}\right.\nonumber\\
  &\pm&\left.\sqrt{1+2(2n+1) \frac{\epsilon_{\alpha}({\bf k})}{U}+\frac{\epsilon_{\alpha}({\bf k})^2}{U^2}}\right),
  \label{eq:miex}
\end{eqnarray}
where $\alpha$ takes values $\pm$ corresponding to the two non--interacting bands. A detailed study of the properties of excitations given by Eq.~(\ref{eq:miex}) is postponed to Sec. \ref{Sec:EXC}. Here we only note that at filling $n=1$ the gap in the spectrum closes at $t_1^{\mathrm{RPA}, c} = U\frac{3-2\sqrt{2}}{3}$ for $t_1>\sqrt{3}t_2$ where the transition from the Mott insulator into the uniform superfluid occurs, while for $t_1<\sqrt{3}t_2$ the transition into the  chiral superfluid is found for $t_2^{\mathrm{RPA}, c} = U \frac{3-2\sqrt{2}}{3\sqrt{3}}$. The two boundaries meet at the tricritical point with the line corresponding to a direct transition between the two superfluids, Fig.~\ref{Fig:Fig1}b.

Another important feature of the Mott state are finite (non--vanishing) density fluctuations present at finite values of the hopping terms $t_1$ and $t_2$ (note that these are not captured by the oversimplified state $\prod_{\otimes i} |n\rangle$). To understand how these fluctuations are affected by the 
complex hopping term ($i t_2$), we calculate local (plaquette) currents (\ref{eq:jij})  between next--nearest neighbors. 
 The expectation value $\langle b_i^\dagger b_j\rangle $ can be expressed in terms of Green's functions
\begin{equation}
\langle b_i^\dagger b_j\rangle=-\frac{1}{\beta} \sum_n \exp(i \omega_n 0^+) G_{ji}(i\omega_n) ,
\end{equation}
where $\omega_n$ is Matsubara frequency. 
Deep in the Mott domain, an approximate result for $G_{ij}$ can be derived from the strong coupling expansion in hopping \cite{Freericks1, Freericks2}. Here we consider only contributions obtained by a formal matrix inversion of Eq.~(\ref{eq:realspacedysonsc}) and by keeping terms that are second order in $t_{ij}$. Directly from the second order result for $\langle b_i^\dagger b_j\rangle $ given in Refs.~\cite{Freericks1, Freericks2}, we read off a general expression
\begin{equation*}
\frac{J_{ij}^{(2)}}{U}=-2 \mathrm{Im}\, t_{ij}\left(\sum_k t_{jk}t_{ki}\right)\frac{3 n (n+1) (2 n +1)}{U^3}+\ldots
\end{equation*}
In our case, we obtain the following perturbative result:
\begin{equation}
 \frac{J_{AA}^{(2)}}{U}=\frac{36}{U^3} t_2\left(t_1^2-2 t_2^2\right).
 \label{eq:secordcurr}
\end{equation}
On the other hand, at this order in perturbation theory the current between nearest neighbors vanishes. 

We finally note that correlations $\langle J_{AA} J_{AA} \rangle$ between currents belonging to plaquettes which are separated [\textit{i.e.} not connected by a kinetic term in $\mathcal{H}_\text{H}$ of Eq.~(\ref{eq:hamk})] factorize and are proportional to the square of Eq.~(\ref{eq:secordcurr}). Therefore, the connected correlation functions of plaquette currents vanish identically for separated plaquettes. This characterizes the Mott insulator phase as a state with local plaquette currents \textit{without long range current-current correlations}.

As mentioned at the beginning of the subsection, RPA is an exact description in the limit of infinite coordination number of the lattice. In the next section we will use bosonic DMFT to include the next--order correction (\textit{i.e.}~a finite coordination number) and to go beyond RPA.

\subsection{DMFT}
\label{Subsec:DMFT}
Bosonic DMFT was originally introduced several years ago \cite{bdmft1, bdmft2, Hubener, Snoek&Hofstetter, Anders}
in analogy to the well--established fermionic DMFT \cite{FermionicDMFT}. The method has been successfully applied in the context of topological band insulators
with fermions \cite{WuRachel, Cocks, OrthCocks, AraGo}.
Here we use a spatially resolved version, the so--called real--space bosonic DMFT \cite{LiHof}.
For completeness, we describe the method briefly.

The essence of DMFT is mapping of the full lattice problem onto a set of local problems. 
The next order correction for the self--energy in  Eq.~(\ref{eq:rpaselfenergy}) is still local in space
 [in our case proportional to $3 (t_1^2+2 t_2^2)$] as can be shown by a diagrammatic expansion \cite{Freericks1}.
To derive a proper local model that goes beyond the  Hamiltonian (\ref{eq:localmfh}), we perform an effective integration over all off--site degrees of freedom and keep only terms of suitable order. We find that the local Hamiltonian is given by a bosonic Anderson impurity model
\begin{eqnarray}
 \mathcal{H}_\mathrm {AI}^i &=& \sum_{l=0}^L \left[\varepsilon_l \hat a_l^{\dagger} \hat a_l + V_l \hat a_l^{\dagger} \hat b_i + V_l^* \hat a_l \hat b_i^{\dagger} +  W_l \hat a_l \hat b_i + W_l^* \hat a_l^{\dagger} \hat b_i^{\dagger}\right]\nonumber\\
&-&\psi_i^{\mathrm {AI}*} \hat b_i - \psi_i^{\mathrm {AI}} \hat b^{\dagger}_i + \frac{U}{2} \hat n_i (\hat n_i-1)-\mu \hat n_i,
\label{eq:Anderson}
\end{eqnarray}
where index $l$ counts Anderson orbitals  and we allow for complex values of Anderson parameters $V_l$ and $W_l$. The mapping is formally described in detail in Ref.~\cite{Snoek&Hofstetter}. At this point it is useful to introduce hybridization functions of the Anderson impurity model
\begin{eqnarray}
 \Delta_{11}(i\omega_n) &=& \sum_l \frac{|V_l|^2}{\varepsilon_l -i \omega_n}+\frac{|W_l|^2}{\varepsilon_l +i \omega_n},\nonumber\\
 \Delta_{12}(i\omega_n) &=& \sum_l \frac{V_l^*W_l^*}{\varepsilon_l-i \omega_n}+\frac{V_l^* W_l^*}{\varepsilon_l+i\omega_n},
 \label{eq:hyb}
\end{eqnarray}
and $\Delta_{21}(i\omega_n)=\Delta_{12}(i\omega_n)^*$, $\Delta_{11}(i\omega_n)=\Delta_{22}(i\omega_n)^*$.
The term $\psi_i^{\mathrm{AI}}$ used in Eq.~(\ref{eq:Anderson}) incorporates a correction with respect to the mean--field result and it reads \cite{Snoek&Hofstetter}:
\[
 \psi_i^{\mathrm{ AI}} =  \sum_{j} t_{ij} \psi_j  - \Delta_{11}(0)\psi_i - \Delta_{12}(0)\psi_i^*.
\]
We consider local Green's functions written in the Nambu notation to take into account off--diagonal terms
\[
 G_{ii}(\tau_1-\tau_2) = -\left(\begin{array}{cc}
                                \langle \mathcal{T} \hat b_i(\tau_1) \hat b_i^{\dagger}(\tau_2)\rangle & \langle \mathcal{T} \hat b_i(\tau_1) \hat b_i(\tau_2) \rangle\\
                                \langle \mathcal{T} \hat b^{\dagger}_i(\tau_1) \hat b^{\dagger}_i(\tau_2)\rangle & \langle \mathcal{T} \hat b_i^{\dagger}(\tau_1) \hat b_i(\tau_2) \rangle
                                \end{array}
                          \right).  
\]
The self--energy is obtained from the local Dyson equation 
\begin{equation}
 G^{-1}_{ii} (i \omega_n) =  \left( \begin{array}{cc}
                          i \omega_n +\mu + \Delta_{11} - \Sigma_{i}^{11}& \Delta_{12}-\Sigma_{i}^{12} \\
                         \Delta_{21}- \Sigma_{i}^{21} \ & -i \omega_n + \mu +\Delta_{22} - \Sigma_{i}^{22}
                         \end{array}\right).
 \label{eq:localdysonequation}
\end{equation}
In analogy to Eq.~(\ref{eq:realspacedysonsc}), the real-space Dyson equation takes the following form:
\begin{equation}
 G^{-1}_{ij, \mathrm{latt}} (i \omega_n) =  \!\left( \begin{array}{cc}
                          \!\!\left(i \omega_n \!+\!\mu\!  -\! \Sigma_i^{11}\right)\delta_{ij}\!+ \!t_{ij}& \!\!-\Sigma_i^{12} \delta_{ij}\\
                         \!\!- \Sigma_i^{21} \delta_{ij}\ & \!\!\left(\!-\!i \omega_n \!+\! \mu \!-\! \Sigma_i^{22}\right)\delta_{ij}\!+\!t^*_{ij}\!\! 
                         \end{array}\right),
                         \label{eq:realspacedyson}
\end{equation}
where we approximate the self--energy by a local contribution from Eq.~(\ref{eq:localdysonequation}).
Finally, we need a criterion to set values of parameters $\varepsilon_l$, $V_l$ and $W_l$ in Eq.~(\ref{eq:Anderson}).  To this end, a condition is imposed on the hybridization functions (\ref{eq:hyb}).  These functions should be optimized such that the two Dyson equations, (\ref{eq:localdysonequation}) and  (\ref{eq:realspacedyson}), yield the same values of local Green's functions. Therefore, local correlations are treated beyond the mean--field level.

In practice, we iterate a self--consistency loop to fulfil this condition, starting from arbitrary initial values. The local problem (\ref{eq:Anderson}) is solved by exact diagonalization and we obtain results for the local density $n_i=\langle \hat n_i\rangle$, density fluctuations $\Delta n_i = \langle \hat n_i^2\rangle-\langle \hat n_i\rangle^2 $ and local condensate order parameter $\psi_i =\langle \hat b_i\rangle $. Here we work with a finite lattice consisting of 72 sites that provides the proper sampling of the Brillouin zone that includes its corners \cite{Varney}. To benchmark our code, we compare our results for hexagonal and triangular lattice (without flux) with accurate results available in the literature \cite{Teichmann}. The deviation in the position of tip of the first Mott lobe is of the order of several percent and it is smaller for triangular than for  hexagonal lattice, which can be justified by a higher coordination number of the former lattice. 

\begin{figure}[!hbt]
\includegraphics[width=\linewidth]{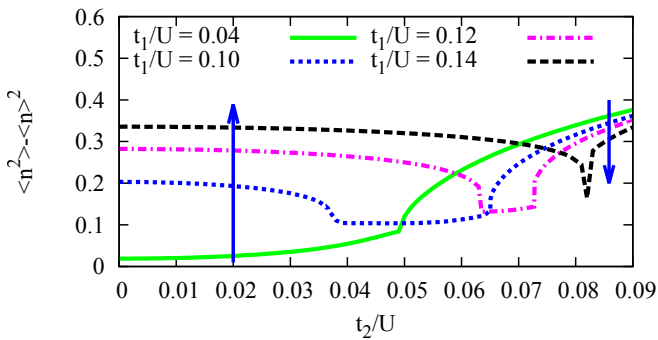}
\caption{(Color online) Local density fluctuations  $\langle \hat n_i^2\rangle-\langle \hat n_i\rangle^2 $ vs. $t_2$ at unit filling $n=1$. We see a competing effect of $t_1$ and $t_2$, as indicated by arrows: for small values of $t_2$, $t_1$ enhances fluctuations and drives the transition into SF. The opposite effect is found for strong enough $t_2$ in \textit{CSF}, where eventually the effect of $t_1$ is washed out.}
\label{Fig:Figdfl}
\end{figure}

\begin{figure}[!tbh]
\begin{center}
\includegraphics[width=\linewidth]{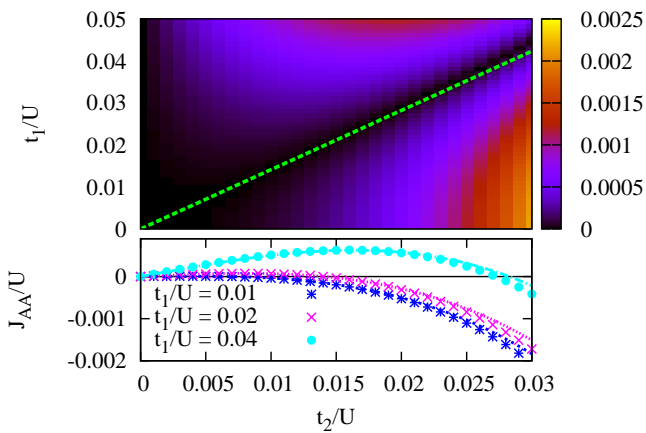}
  \end{center}
\caption{(Color online) Top: Absolute value of the next--nearest neighbor bond current  $|J_{A A}|/U $ deep in MI. The dashed line $t_1=\sqrt{2} t_2$ marks the region $J_{AA}=0$  according to Eq.~({\ref{eq:secordcurr}}). Bottom: DMFT data for $J_{A A}$ (dots) agree very well with the result Eq.~({\ref{eq:secordcurr}}) (lines) in this region of the phase diagram.}
\label{Fig:FigPMIC}
\end{figure}

\begin{figure}[!bt]
\includegraphics[width=\linewidth]{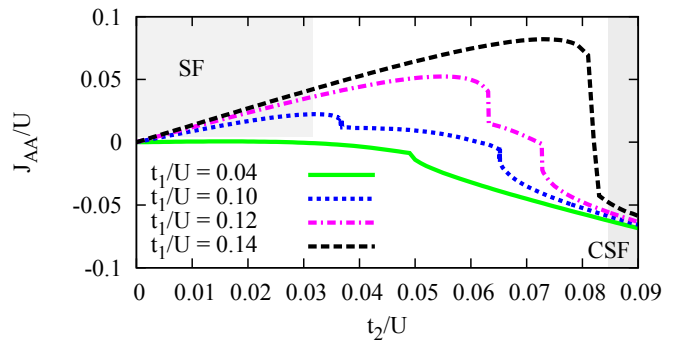}
\caption{(Color online) The $J_{AA}$ current vs. $t_2$  for $n=1$ and several values of $t_1$. Deep in SF (upper left) the current is positive and exhibits linear increase with $t_2$.  In CSF (bottom right) the current is negative, strongly dependent on $t_2$ and only weakly affected by $t_1$. The absolute value $|J_{AA}|$ is much weaker in MI (intermediate regions).}
\label{Fig:Figcurr}
\end{figure}

The resulting phase diagram for the model (\ref{eq:hami}) is given in Fig.~\ref{Fig:Fig1}b. 
To have $n=1$ filling on the superfluid side we adjust the value of the chemical potential $\mu$.
We find that the Mott domain is extended in comparison to the mean--field result
and its ``cusp'' shape reveals the subtle interplay of two types of hopping, which goes beyond the simple mean--field picture. We can understand this point better by looking at density fluctuations, Fig.~\ref{Fig:Figdfl} and plaquette currents, Fig.~\ref{Fig:FigPMIC} and Fig.~\ref{Fig:Figcurr}.

In Fig.~\ref{Fig:Figdfl}, we show local density fluctuations for several cuts through the phase diagram.
Weak density fluctuations are a hallmark of the Mott insulator phase, while stronger density
fluctuations correspond to superfluid phases.  In the \textit{CSF} phase (rightmost part of the plot) $t_1$ supresses density fluctuations. 
This relates to the fact that $t_1$ pushes the \textit{PMI}-\textit{CSF} phase boundary  toward higher values of $t_2$, Fig.~\ref{Fig:Fig1}b.
Deep in the \textit{CSF} phase ($t_2/U>0.08$)  the effect is very weak (different curves become indistinguishable) in accordance with our mean--field results. At strong enough values of $t_1$ we enter the SF phase (curves in the upper left part of the plot). Finally, dips in the curves  mark the reentrant transition from \textit{SF} into \textit{PMI}.

Next, we turn to bond currents between next--nearest neighbors, Fig.~\ref{Fig:FigPMIC} and Fig.~\ref{Fig:Figcurr}. 
In Fig.~\ref{Fig:FigPMIC} we observe that in the \textit{PMI} our numerical results are in good agreement with Eq.~(\ref{eq:secordcurr}) and the current $J_{AA}$ changes its sign smoothly here. In the limit of weak hopping the sign change occurs for $t_1 = \sqrt{2} t_2$, Fig.~\ref{Fig:FigPMIC}. 
At stronger values of $t_1$ and $t_2$, there are more features showing up, Fig.~\ref{Fig:Figcurr}.
In agreement with Eqs.~(\ref{eq:current1}) and (\ref{eq:current2}), $J_{AA}$ is positive in the \textit{SF}, and negative in the \textit{CSF}.
By increasing $t_2$ and keeping $t_1$ small enough (for example $t_1/U = 0.04$ in Fig.~\ref{Fig:Figcurr}) we reach the point of the second order \textit{PMI}--\textit{CSF} transition.
The aforementioned reentrant phase transition is marked by two second--order phase transitions 
from the \textit{SF} into \textit{PMI} and from the \textit{PMI} into the \textit{CSF}, for example for $t_1/U=0.10$. It is interesting to note the nonmonotonic behavior - the absolute value of $J_{AA}$ initially exhibits a linear increase with $t_2$, but as we approach the Mott domain it decays due to a reduced value of the order parameter. At strong enough  $t_1$ and $t_2$, we expect the intermediate Mott domain to vanish and the first order phase transition described in section \ref{sec:SecWIB} to set in.

\begin{figure}[!tbh]
\includegraphics[width=\linewidth]{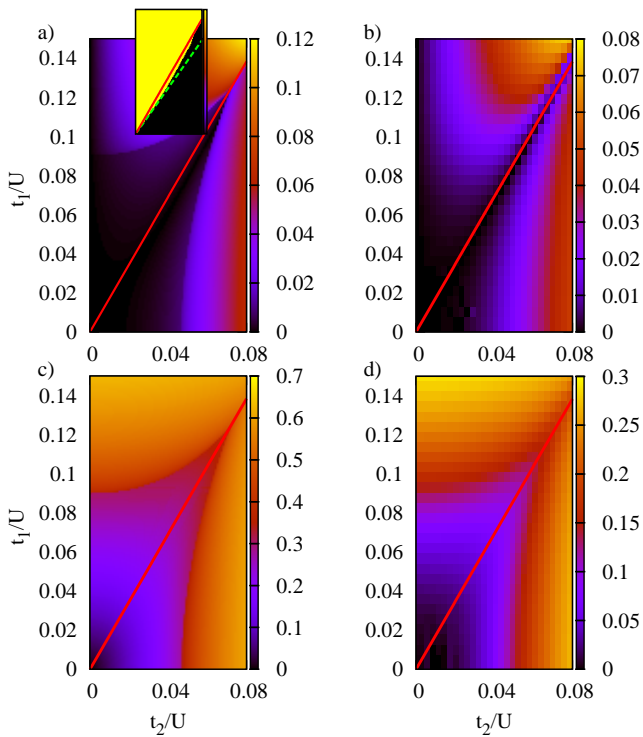}
\caption{(Color online) Comparison of DMFT (a) and c)) and ED results (b) and d)). Absolute value of the current on the bond between two next--nearest neighbors $|J_{AA}|/U$ is shown in a) and b). In c) and d) we plot local density fluctuations  $\langle \hat n_i^2\rangle-\langle \hat n_i\rangle^2 $. 
The inset in a) gives the sign of $J_{AA}$: In the superfluid, the sign changes for $t_1=\sqrt{3}t_2$ (solid line), while deep in the Mott domain the boundary is given by $t_1=\sqrt{2} t_2$ (dashed line). DMFT data shown in these plots are for the fixed value of chemical potential $\mu/U = 0.4$, while ED results are at fixed filling $n=1$. However, the main features discussed in the text are clearly visible.}
\label{Fig:Figcomp}
\end{figure}

\begin{figure}[!tbh]
\begin{center}
\includegraphics[width=\linewidth]{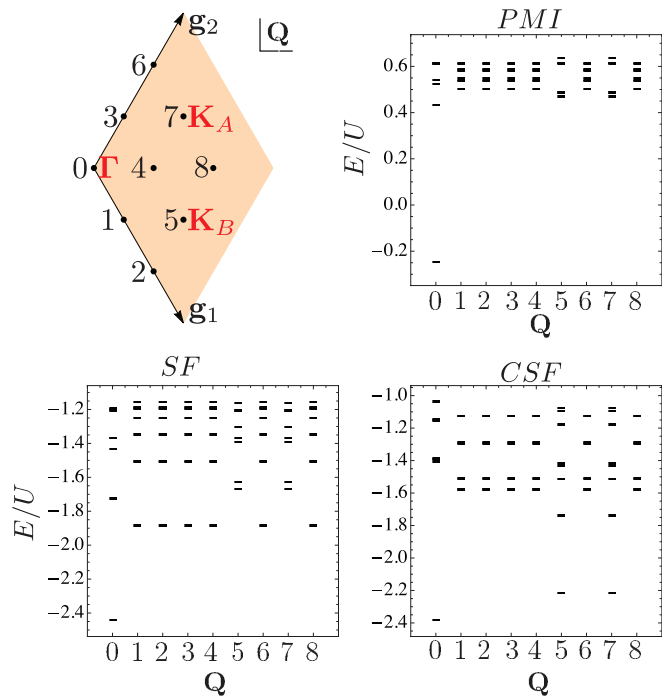}
\end{center}
\caption{(Color online) Low lying energy levels in the spectra of the $3 \times 3$ unit cell lattice at unit filling, for: \textit{PMI} ($t_1/U = 0.04$, $t_2/U = 0.02$), \textit{SF} ($t_1/U = 0.14$, $t_2/U = 0.02$), and \textit{CSF} ($t_1/U = 0.04$, $t_2 = 0.08$). The first 8 energy levels in each sector of total momentum $\mathbf{Q}$ are plotted. The 9 momentum sectors in the Brillouin zone spanned by the vectors $\mathbf{g}_1$ and $\mathbf{g}_2$ are represented in the top left panel. For example, sectors labeled 0, 7, 5 correspond to $\mathbf{\Gamma}$, $\mathbf{K}_A$ and $\mathbf{K}_B$, respectively. }
\label{Fig:SpectraED}
\end{figure}

\subsection{Exact diagonalization}
\label{Subsec:Lanczos}
In this section we use the ALPS implementation \cite{ALPS} of the Lanczos algorithm \cite{CullumWilloughby} in order to study the ground state of the interacting model in Eq.~(\ref{eq:hami}) at unit filling $n = 1$. We consider a lattice of $3 \times 3$ unit cells, implying $N_\textit{sites} = 18$ and $N=18$ particles. The truncated boson Hilbert space contains states for which the number expectation value at any site is bounded above $\langle n_i \rangle \leq 2$. The Hilbert subspace with this constraint has dimension 44152809. With periodic boundary conditions, total momentum $\mathbf{Q} = \sum_{i=1}^{N} \mathbf{k}_i$ is a good quantum number and the dimension of the Hilbert space for each momentum sector is reduced by a factor of $9$. The Brillouin zone contains $3 \times 3$ points and includes the inequivalent points $\mathbf{\Gamma}$ and $\textbf{K}_A, \textbf{K}_B$, as shown in Figure~\ref{Fig:SpectraED}.

Since total particle number is conserved, the spontaneous breaking of the $U(1)$ symmetry is not observable in the ground state. We rather identify the Mott insulator phase as the region of the $\left( t_1 / U, t_2 / U \right)$ plane where number fluctuations at a site $\langle n_i^2 \rangle - \langle n_i \rangle^2$ are small [see Figure~\ref{Fig:Figcomp}d)].

As shown in Sec.~\ref{sec:SecWIB}, bond current expectation values distinguish the chiral superfluid from the uniform superfluid phase. The nearest neighbor current $J_{AB}$ vanishes identically at $n = 1$. The next-nearest neighbor bond current $J_{AA}$ as a function of $t_{1}$ and $t_{2}$ is consistent with the result from strong-coupling perturbation theory [Eq.~(\ref{eq:secordcurr})]. The next-nearest neighbor current $J_{AA}$ changes sign at $t_1 = t_2 \, \sqrt{3}$, and $J_{BB}$ has analogous behavior [their common absolute value is plotted in Figure~\ref{Fig:Figcomp}b)]. This is the exact phase boundary found previously in the weakly interacting regime and with DMFT for arbitrarily strong interactions. We thus confirm the existence of the \textit{PMI} state with nonzero triangular plaquette currents at order $t_1^2 t_2/ U^3$ but vanishing nearest-neighbor currents. 

To understand the momentum structure of the ground state, we consider the momentum distributions,
\begin{equation}
n_{a} (\mathbf{k}) \equiv \langle \hat{b}^\dagger_{a}(\mathbf{k}) \hat{b}_{a}(\mathbf{k}) \rangle = \sum_j e^{ i  \mathbf{k} \cdot \mathbf{r}_j } \langle \hat{b}^\dagger_{a 0}   \hat{b}_{a j}\rangle,
\end{equation}
where $a = A$ or $B$ denotes the sublattice. In agreement with analytical results in the weakly interacting limit [Sec.~\ref{sec:SecWIB}], we find that in the \textit{SF} the momentum distribution $n_{A}(\mathbf{k}) + n_{B}(\mathbf{k})$ is sharply peaked at the $\mathbf{\Gamma}$ point. In the \textit{CSF} phase, $n_{a}(\mathbf{k})$ are peaked at $\mathbf{K}_{a}$, in agreement with the decoupled condensates wavefunction $|\Phi_0''\rangle$ of Eq.~(\ref{eq:gs2}). In the \textit{PMI}, $n_{a}(\mathbf{k})$ become more and more uniformly distributed as the hopping amplitudes $t_1$ and $t_2$ approach 0. Regarding the symmetries, we remark that the 9 lattice translation symmetry operators and the lattice inversion symmetry are conserved by the ground state in the $\mathbf{Q}=0$ sector. Without breaking the discrete lattice symmetries it is impossible to obtain the coherent superposition $|\Phi_0'\rangle$.

In conclusion, we confirm by studying the exact ground state of the $3\times 3$ lattice all the qualitative features of the DMFT phase diagram [see Figure~\ref{Fig:Figcomp} for a comparison]. We can distinguish between the \textit{SF} and the \textit{CSF} by studying momentum distributions and currents, and determine sharply the phase boundary between the two superfluids at the critical line $t_1 = t_2 \sqrt{3}$ by detecting the change in sign of the next-nearest neighbor currents $J_{AA}$ or $J_{BB}$. 

 We stress that the results obtained in this section are for a finite sized lattice Hamiltonian obeying translation and inversion symmetries. Consequently, the finite size ground state in the $\mathbf{Q}=0$ sector obeys these symmetries. For an infinite system obeying all symmetries, the ground state may be identified as a specific linear combination of degenerate ground states that breaks the symmetries \cite{Moller}. To briefly explore this possibility, we plot the low energy spectra for each phase, as shown in Figure~\ref{Fig:SpectraED}. In the \textit{CSF} phase, there are two low-lying states in the $\mathbf{Q}=\mathbf{K}_A$ and $\mathbf{Q}=\mathbf{K}_B$ sectors, which may become degenerate with the ground state for an infinite system or as $U \to 0$. However, an analysis of the scaling of the gap with system size is limited by the large dimension of the Hilbert space at unit filling. Moreover, the $U \to 0$ limit cannot be rigorously explored numerically due to the necessary truncation of the bosonic Hilbert space.

\section{Excitations}
\label{Sec:EXC}
In this section we study the excitation spectra in the superfluid phases (Subsec. \ref{Subsec:EXCSF}) and in the Mott insulator (Subsec. \ref{Subsec:EXCMI}).

\subsection{Weakly interacting bosons}
\label{Subsec:EXCSF}

To compute the excitation spectrum in the weakly interacting limit, we start from the time--dependent GP equation  \cite{GPbook}
\begin{equation}
 i \frac{\partial \psi_i}{\partial t}=-t_1\sum_{\langle i| j\rangle}\psi_j-it_2\sum_{\langle \langle i| j\rangle \rangle} (\pm )\psi_j+U|\psi_i|^2 \psi_i
\end{equation}
where we make explicit with $\langle i| j\rangle$ and $\langle \langle i| j\rangle \rangle $ that index $i$ is now fixed we are summing over its neighbors. The $\pm$  refers to the sign of the imaginary hopping term along the different directions (additionally, it takes opposite values for the two sublattices).
We expand the order parameter in terms of fluctuations $\delta_i$ around the mean--field solution $\psi_i^0$ as
\begin{equation}
 \psi_i=\left(\psi_i^0+\delta_i\right)\exp\left(-i\mu t\right).
\end{equation}
In the zeroth order in $\delta_i$ we recover the ground state equation 
\begin{equation}
 \mu \psi_i^0 = -t_1\sum_{\langle i |j\rangle}\psi_j^0-\sum_{\langle \langle i | j \rangle \rangle} i t_2 (\pm)\psi_j^0+U|\psi_i^0|^2 \psi_i^0,
 \label{eq:gs}
\end{equation}
that corresponds to the ground state energy given by Eq.~(\ref{eq:gpenergy}).
By keeping terms of the order $\delta_i$ we obtain an equation that allows the study of excitations:
\begin{eqnarray}
 i\frac{\partial \delta_i}{\partial t}&=&-\mu\delta_i-t_1\sum_{\langle i |j \rangle}\delta_j -i t_2\sum_{\langle \langle i |j\rangle \rangle}(\pm )\delta_j\nonumber\\&+&U\left(2|\psi_i^0|^2 \delta_i+(\psi_i^0)^2 \delta_i^*\right).
 \label{eq:eqdel}
\end{eqnarray}

To decouple Eq.~(\ref{eq:eqdel}) further, we proceed in the standard way:
\begin{equation}
 \delta_i=u_i \exp(-i \omega t)+v_i^* \exp(i \omega t),
\end{equation}
and obtain the set of equations:
\begin{eqnarray}
 \omega u_i^A &=& \delta u_i^A-t_1 \sum_{\langle i| j\rangle}\!u_j^B-i t_2 \!\!\sum_{\langle \langle i |j\rangle \rangle}\!\!u_j^A (\pm )+U   (\psi_i^0)^2  v_i^A,\nonumber\\
 -\omega v_i^A &=& \delta v_i^{A}-t_1 \sum_{\langle i| j\rangle}\!v_j^B-it_2 \!\!\sum_{\langle \langle i |j\rangle \rangle}\!\!v_j^A (\mp )+U ((\psi_i^0)^2)^*  u_i^A,\nonumber\\
  \omega u_i^B &=& \delta u_i^B-t_1 \sum_{\langle i |j\rangle}\!u_j^A-it_2 \!\!\sum_{\langle \langle i |j\rangle \rangle}\!\!u_j^B (\mp )+U (\psi_i^0)^2 v_i^B,\nonumber\\
 -\omega v_i^B &=& \delta v_i^{B}-t_1 \sum_{\langle i |j\rangle}\!v_j^A-it_2\! \!\sum_{\langle \langle i |j\rangle \rangle}\!\!v_j^B (\pm )+U  ((\psi_i^0)^2)^* u_i^B, \nonumber\\ \;
 \label{eq:eqset}
\end{eqnarray}
where $\delta = 2 U n-\mu$. In the next step, we rewrite Eq.~(\ref{eq:eqset}) in the momentum basis and solve the emerging eigenproblem.

\begin{figure}[!tbh]
\begin{center}
\includegraphics[width=\linewidth]{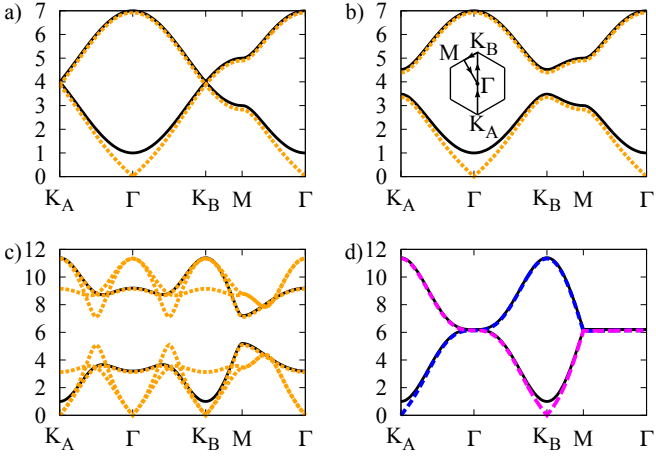}
\end{center}
\caption{(Color online) Bogoliubov dispersion relation $\omega({\bf k})/U$ for $n U =1$ and  a) $t_1=1, t_2=0$, b) $t_1=1, t_2=0.1$, c) $t_1=1, t_2=1$, d) $t_1=0, t_2=1$. Black solid lines give the corresponding non--interacting dispersion relation shifted upwards by $2Un-\mu$. The inset in b) defines the path in the Brillouin zone.}
\label{Fig:Figbg}
\end{figure}

For the \textit{SF} phase, the relevant matrix is of size $4\times 4$, and since Eqs.~(\ref{eq:eqset}) exhibit particle--hole symmetry we obtain two particle excitation branches, shown in Fig.~\ref{Fig:Figbg}a and Fig.~\ref{Fig:Figbg}b. For the honeycomb lattice without flux, Fig.~\ref{Fig:Figbg}a, we find a Goldstone mode and a Dirac cone inherited from the non--interacting dispersion relation, shown in the same figure by a solid line. As we increase $t_2$ but remain in the same phase, the Goldstone mode persists and the sound velocity is  unaffected by $ t_2$.  
The gap between the two bands in Fig.~\ref{Fig:Figbg}b implies that the edge mode structure present in the non--interacting spectrum, arising due to a non--trivial topological index (\ref{eq:ti}), remains intact by the presence of weak interactions \cite{Petrescu0}. 

In \textit{CSF}, the mean-field order parameter takes the  form:
\begin{equation}
 \psi_i^0=\begin{cases}
         \sqrt{n} \exp(-i {\bf K}_A {\bf r}_i) \exp{i \phi_A}, & i\in A\\
         \sqrt{n} \exp(-i {\bf K}_B {\bf r}_i) \exp{i \phi_B}, & i\in B
        \end{cases},
\end{equation}
where we explicitly consider at the mean-field level two independent superfluids by introducing two angles $\phi_A$ and $\phi_B$. The terms proportional $(\psi_i^{0})^2$ couple different momenta in Eq.~(\ref{eq:eqset})~\cite{Powell1, Powell2}. In the case of $t_1=0$, $u^A_{\bf k}$ couples to $v^A_{{\bf k}-2 {\bf K}_A}$ and $u^B_{\bf k}$ couples to $v^B_{{\bf k}-2 {\bf K}_B}$. (Vectors $-{\bf K}_A$, ${\bf K}_B$
and $2{\bf K}_A$ are equal up to reciprocal lattice vectors). Therefore we obtain two decoupled eigenproblems for
\begin{equation}
 h^A({\bf k}) = \begin{pmatrix}
  \delta -d_3({\bf k}) & n U\\
  -n U &-\delta -d_3({\bf k}+{\bf K}_A)
 \end{pmatrix},
\end{equation}
and 
\begin{equation}
 h^B({\bf k}) = \begin{pmatrix}
  \delta +d_3({\bf k}) & n U\\
  -n U &-\delta +d_3({\bf k}-{\bf K}_A)
 \end{pmatrix},
\end{equation}
where ${\bf d}({\bf k})$ is given in Eq.~(\ref{eq:d}),
that yield the following particle excitations:
\begin{eqnarray}
 \omega_{A}({\bf k}) &=& \frac{1}{2}\left(-d_3({\bf k}) - d_3({\bf k}+ {\bf K}_A)\right.\nonumber\\
 &+&\left.\sqrt{(d_3({\bf k})-d_3({\bf k}+ {\bf K}_A)-2 \delta)^2 -4 n^2 U^2}\right),\\
 \omega_{B}({\bf k}) &=& \frac{1}{2}\left(d_3({\bf k}) + d_3({\bf k}- {\bf K}_A)\right.\nonumber\\&+&\left.\sqrt{(d_3({\bf k})-d_3({\bf k}- {\bf K}_A)+2 \delta)^2-4 n^2 U^2}\right).
\end{eqnarray}

These are shown in Fig.~\ref{Fig:Figbg}d. We observe two Goldstone modes - one corresponiding to the superfluid of sublattice A located around ${\bf K}_A$ and the second one at ${\bf K}_B$. The inversion symmetry, present  when both sublattices are taken into account, provides the particle--hole symmetry also in this case \cite{Powell2}. 

As the $t_1$ term is turned on, the three momenta ${\bf k}$, ${\bf k}- {\bf K}_A $ and ${\bf k}+ {\bf K}_A$ are coupled and the eigenproblem corresponding to Eq.~(\ref{eq:eqset}) is of size $12 \times 12$. The related matrix is explicitly given in the Appendix and it incorporates the angles $\phi_A$ and $\phi_B$. An example of the result for the six particle bands is illustrated in Fig.~\ref{Fig:Figbg}c for $\phi_A=\phi_B$. We observe that the two Goldstone modes persist even at finite $t_1$. Some redundancies are present in Fig.~\ref{Fig:Figbg}c --e.~g.~${\bf K}_A$, ${\bf \Gamma}$, and ${\bf K}_B$ correspond to the same point, since we have not reduced the Brillouin zone in accordance with the coupling between different momenta.

\begin{figure}[!tbh]
\begin{center}
\includegraphics[width=\linewidth]{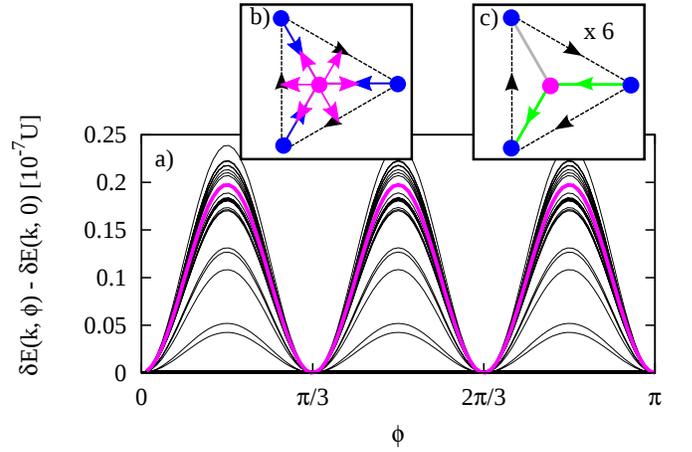}
\end{center}
\caption{(Color online) a) Contribution of quantum fluctuations to the ground state energy $\delta E({\bf k},\phi)-\delta E({\bf k}, 0)$, $t_1=t_2=1$, $n U=1$, $\phi=\phi_A-\phi_B$. For a single value of ${\bf k}$ we obtain the thick curve. The qualitative behavior of  $\delta E({\bf k}, \phi) $ is similar throughout the Brillouin zone (thin lines).  In b) we illustrate the six configurations of order parameters on the two sublattices favoured by quantum fluctuations. In c) the corresponding emerging plaquette currents are depicted.}
\label{Fig:Figqf}
\end{figure}
As mentioned many times throughout the paper, in the \textit{CSF} at the mean--field level the two sublattices are fully decoupled.
We now discuss the role of quantum fluctuations as a beyond mean--field effect that could lift the degeneracy of the mean--field solution and set the value of the phase difference $\phi = \phi_A-\phi_B$.  The mechanism is known as ``order by disorder'' \cite{Henley} and it has been discussed in the context of cold atoms \cite{GPbook2, Barnett, CSSF, Zhai}. The zero--point energy is given by
\begin{equation}
 \delta E(\phi) = \sum_{{\bf k}}\delta E({\bf k}, \phi) = \frac{1}{2}\sum_{{\bf k}, l} \omega_l({\bf k}, \phi)
\end{equation}
as can be derived using the Bogoliubov approach in the operator formalism \cite{GPbook, Barnett, Zhai}. The index $l$ enumerates six particle bands. In Fig.~\ref{Fig:Figqf}a, we show the typical behavior of $\delta E({\bf k},\phi)-\delta E({\bf k}, 0)$ throughout the Brillouin zone. The energy differences are typically small, but configurations with $\phi = m \times \frac{\pi}{3}$, where $m$ is an integer (Fig.~\ref{Fig:Figqf}b) are preferred. Phase ordering between the sublattices leads to the current patterns shown in Fig.~\ref{Fig:Figqf}c. These are similar to the patterns discussed in the subsection  \ref{Subsec:CrtWeak}: out of the three nearest--neighbor links shown, one does not carry current and there are two possible directions for the current flow on the two other links. 
These current patterns are periodic with respect to an enlarged unit cell which consists of six sites. 
The dependence $1-\cos(6\phi)$ can be traced back to the eigenproblem
of the matrix (\ref{eq:hamcsf}). 

By inspecting the characteristic polynomial  (\ref{eq:polynom})
we find that the only contribution of the phase difference $\phi$ is in the free term as $\cos \left(6 \phi\right)$ with proportionality constant $t_1^6 U^6$. The contribution vanishes exactly at ${\bf K}_B$ and ${\bf K}_A$.

Our findings comply to the general rule stating that quantum fluctuations favor colinear (parallel or anti-parallel) order parameters \cite{Henley}. As cold atoms represent highly tunable and clean systems,  the effect could be possibly observed in an experiment. However, it may be difficult to distinguish it  from the pinning that arises due to the boundary conditions or defects, discussed in subsection \ref{Subsec:CrtWeak}. In addition, thermal effects may be important, but we do not discuss this point further.

\subsection{Excitations of the Mott phase}
\label{Subsec:EXCMI}
In this section we study quasiparticle or quasihole excitations of the Mott insulator phase. We use the single particle Green's function $G( i \omega_n, \mathbf{k} )$ to compute quasiparticle and quasihole band dispersions. We characterize transport in excited bands through band Chern numbers \cite{TKNN}. We obtain $G( i \omega_n, \mathbf{k} )$ from DMFT and from the strong coupling random phase approximation \cite{Senechal,Senechal1, Senechal2}.

\subsubsection{Strong coupling expansion}
We use the results of the strong coupling expansion with RPA introduced in Subsec.~\ref{Subsec:Gutzwiller} to study the spectrum of quasiparticle and quasihole excitations. We extend the approach of Subsec.~\ref{Subsec:Gutzwiller} by grouping the sites on the lattice into identically shaped nonoverlapping clusters \cite{Senechal,Senechal1,Senechal2} (\textit{e.g.} the collection of unit cells pointing along $\mathbf{a}_1$ is a collection of 2 site clusters). Starting from the limit of decoupled clusters (intercluster hopping vanishes) we treat intercluster hopping perturbatively, summing all RPA contributions.

Let $\mathcal{H}_\text{H}'$ be the sum of intercluster hopping terms in $\mathcal{H}_\text{H}$. Let $\mathcal{H}_\text{I} \equiv \mathcal{H} - \mathcal{H}_\text{H}$ denote the interaction part of the Hamiltonian. The Hamiltonian of decoupled clusters is
\begin{equation}
\mathcal{H}_\text{C} = \mathcal{H}_\text{I} + \mathcal{H}_\text{H} - \mathcal{H}_\text{H}' = \sum_j \mathcal{H}_{\text{C}j}.
\end{equation}
The sum in the second equality is over decoupled cluster Hamiltonians $\mathcal{H}_{\text{C}j}$.

We now define the \textit{local} Green's function corresponding to one decoupled cluster. Let the ground state of $\mathcal{H}_{\text{C}j}$ be $| \Phi_{0j} \rangle$ with ground state energy $E_{0}$. Denote sites within a cluster using Latin indices $a,b = A$ or $B$, such that $\hat{b}_{aj}$ annihilates a quasiparticle at the $a^\text{th}$ site of the $j^\text{th}$ cluster. The local Green's function is
\begin{eqnarray}
\label{Eq:grpa}
[ G^\mathrm{RPA}_{jj} ( i \omega_n ) ]_{ab} &=& - \langle \Phi_{0j} | \hat{b}^\dagger_{bj} \frac{1}{i\omega_n - E_0 + \mathcal{H}_{\text{C}j} } \hat{b}_{aj} | \Phi_{0j} \rangle \nonumber \\ 
&& + \langle \Phi_{0j} | \hat{b}_{aj} \frac{1}{i\omega_n + E_0 - \mathcal{H}_{\text{C}j}} \hat{b}_{bj}^\dagger | \Phi_{0j} \rangle,
\end{eqnarray}
for each cluster $j$. In what follows, we assume that clusters are identical, and therefore we will drop the cluster index denoting the local Green's function simply by $[G^{\mathrm{RPA}}(i \omega_n )]_{ab}$.

Note that Eq.~(\ref{Eq:grpa}) reduces to Eq.~(\ref{eq:gii}) of Subsec.~\ref{Subsec:Gutzwiller} if we consider single-site clusters. The spectral function has a pole at $n U - \mu$ with residue $(n + 1)$ and a pole at $(n-1)U - \mu$ with residue $-n$. If the cluster comprises the unit cell, hybridization from the intracluster kinetic term results in pairs of quasiparticle and quasihole poles.

\begin{figure}[t!]
\includegraphics[width=\linewidth]{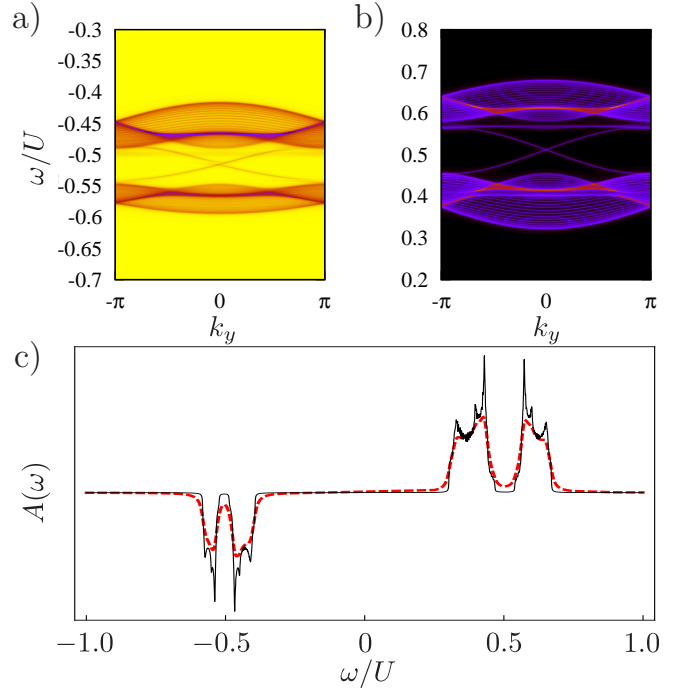}
\caption{\label{Fig:FigEXC1} (Color online) Typical spectral functions deep in the Mott domain, for $t_1 = 3U/100$, $t_2 = U/100$, $\mu = U/2$: \textbf{a)} and \textbf{b)} quasihole and quasiparticle branches of the spectral function obtained from DMFT for a cylinder geometry exhibiting edge modes for $\delta = U/1000$. \textbf{c)} density of states $A(\omega)$ (arbitrary units) in a torus geometry for two values of $\delta$: $U/1000$ (solid black line) and $U/100$  (thick dashed red line). Quasiparticle and quasihole bands are centered at $U - \mu$ and $-\mu$, respectively.}
\end{figure}

We commit to clusters consisting of a single unit cell. In this case, the RPA approximation to the single particle Green's function is \cite{Senechal1,Senechal2}
\begin{equation}
\label{Eq:invGSC}
\left[ G^\mathrm{RPA}( i \omega_n, \mathbf{k} ) \right]^{-1} =  [ G^{RPA}( i \omega_n ) ]^{-1} - \mathcal{H}_\text{H}'( \mathbf{ k } ).
\end{equation}
This is the equivalent of Eq.~(\ref{eq:kspacedysonsc}) in the cluster perturbation theory language. The difference is that now all three of $[G^{RPA}( i \omega_n, \mathbf{k} ) ]_{ab}$, $[ G^{RPA}( i \omega_n ) ]_{ab}$ and $[\mathcal{H}_\text{H}'( \mathbf{ k } )]_{ab}$ are $2\times 2$ matrices acting on the sublattice basis. Tracing over sublattice indices in Eq.~(\ref{Eq:invGSC}), we obtain the spectral function 
\begin{equation}
\label{Eq:ASC}
A( \omega, \mathbf{k} )  = - (1/\pi)~\text{Tr}~\text{Im}~G^{\mathrm{RPA}}( \omega, \mathbf{k} ) .
\end{equation}
Since $G^\mathrm{RPA}( \omega, \mathbf{k})$ in Eq.~(\ref{Eq:invGSC}) is a rational function, the spectral function $A( \omega, \mathbf{k} )$ is a sum of Lorentzians. Deep in the Mott phase, the strong coupling spectral function agrees with that obtained from DMFT, plotted in Figure~\ref{Fig:FigEXC1}~a),b) for a finite cylinder geometry. Note that the resolution of the edge states is dependent on the inverse lifetime $\delta$. In Figure~\ref{Fig:FigEXC1}~c) we plot the density of states $A(\omega) \equiv \int d^2\mathbf{k}~A(\omega,\mathbf{k})$ for two values of $\delta$. The gap between quasiparticle (hole) bands disappears when the inverse lifetime approaches the bandwidth, $\delta \sim t_1$. In the opposite regime, a clear gap is present for $\delta \ll t_1$. We note finally that using larger clusters \cite{Senechal1,Senechal2} yields a $G^\mathrm{RPA}$ whose qualitative features are similar to those of Eq.~(\ref{Eq:invGSC}). In particular, this approach will not yield an estimate for the quasiparticle 
lifetime $1/\delta$.

\subsubsection{Chern number of particle or hole excitations}
\label{Subsec:SCChern}
We assume that $\delta \ll t_1$, such that the $G^\mathrm{RPA}( \omega, \mathbf{k} )$ has well defined quasiparticle and quasihole peaks. We use Greek indices $\alpha = +,-$ to denote the upper and lower subbands. We denote the quasiparticle dispersion relation as $\omega_{+,\alpha} ( \mathbf{k} )$ and the quasihole dispersion relation $\omega_{-,\alpha} ( \mathbf{k} )$. Quasiparticle and quasihole poles arise from the equation 
\begin{equation}
\lambda\left[ \omega_{\pm,\alpha}( \mathbf{k} ), \mathbf{k} \right] = 0,
\end{equation}
where $\lambda$ denotes any one of the two eigenvalues of $[G^\mathrm{RPA} (\omega, \mathbf{k})]^{-1}$ obtained from Eq.~(\ref{Eq:invGSC}).  We are interested in band Chern numbers, which arise from the Ishikawa-Matsuyama formula \cite{IshikawaMatsuyama} of the many-body Hall conductivity
\begin{eqnarray}
\label{Eq:N2}
\sigma_{\textit{xy}} = -\int\frac{d^2 \mathbf{k} d\omega}{8\pi^2} \epsilon^{ij} \text{Tr} \left[ \partial_0 G \partial_i G^{-1} G \partial_j G^{-1} \right]. 
\end{eqnarray}
The summation over indices $i, j = 0,1,2$ is implicit and $\epsilon^{ij}$ is the antisymmetric tensor. Integrations are performed over the Brillouin zone and over real frequencies $\omega$. We have denoted partial derivatives as $\partial_j = \partial / \partial k_j$, where $k_0 \equiv \omega$, and $k_{1,2}$ denote momentum.

Let $\mathcal{U}( \omega, \mathbf{k})$ be the unitary transformation that diagonalizes $[G^\mathrm{RPA}( \omega, \mathbf{k} )]^{-1}$, that is
\begin{equation}
[G^{\mathrm{RPA}}]^{-1}_{ab} = \sum_{\alpha\beta} \mathcal{U}_{a\alpha} \lambda_{\alpha} \delta_{\alpha \beta} \mathcal{U}^\dagger_{\beta b}. 
\end{equation}
We introduce the matrix of Berry gauge fields
\begin{equation}
\mathcal{A}_{\alpha \beta}^j = \sum_{a} i\,\mathcal{U}_{a \alpha} \partial_j \mathcal{U}^\dagger_{\beta a}, \text{ for } j = 0,1,2.
\end{equation}
Note that the diagonal component $\mathcal{A}^j_{\alpha \alpha}$ is the Berry gauge field associated with the $\alpha^\text{th}$ band. If the Green's function has only simple poles at $\omega_{\pm,\alpha}(\mathbf{k})$, then the frequency integral of Eq.~(\ref{Eq:N2}) can be performed \cite{WongDuine1, WongDuine2, ShindouBalents}, leading to
\begin{equation}
\label{Eq:N2As}
\sigma_\textit{xy} = - \sum_{\alpha\delta}\int \frac{d^2 \mathbf{k}}{2\pi} \;  \epsilon^{ij} \left[ \mathcal{A}_{\alpha \delta}^i  \mathcal{A}_{\delta \alpha}^j  + v^i_{-,\alpha} \mathcal{A}_{\alpha \delta}^j \mathcal{A}_{\delta \alpha}^0 \right]_{\omega = \omega_{-,\alpha} (\mathbf{ k })}.
\end{equation}
The frequency integral of Eq.~(\ref{Eq:N2}) amounts to evaluating the integrand of Eq.~(\ref{Eq:N2As}) at the two quasihole poles $\omega_{-,\alpha}(\mathbf{k})$. We have introduced band velocities 
\begin{equation}
v^j_{-,\alpha}( \mathbf{k} ) \equiv \partial_j \omega_{-,\alpha}( \mathbf{k} ).
\end{equation}

To further simplify Eq.~(\ref{Eq:N2As}), define the \textit{on-shell} Berry gauge field for quasihole bands as
\begin{equation}
\label{Eq:Bhab}
\mathcal{B}_{h,\alpha \beta}^i ( \mathbf{k} ) = \sum_{a} i\,\mathcal{U}_{a\alpha} \left[ \omega_{-,\alpha} (\mathbf{k}), \mathbf{k} \right] \partial_i \mathcal{U}^\dagger_{\beta a} \left[ \omega_{-,\beta}( \mathbf{k} ), \mathbf{k} \right].
\end{equation}
Then $\sigma_\textit{xy}$ measures the flux of the on-shell Berry field strength through the Brillouin zone and splits into a sum over quasihole bands $\sigma_\textit{xy} = \sum_{\alpha = \pm} \mathcal{C}_\alpha$, where
\begin{eqnarray}
\label{Eq:CalphaLast}
\mathcal{C}_\alpha =  \frac{1}{2\pi} \int d^2 \mathbf{k} \left[ \partial_1 \mathcal{B}_{h,\alpha \alpha}^2( \mathbf{k} ) - \partial_2 \mathcal{B}_{h,\alpha \alpha}^1( \mathbf{k} )  \right].
\end{eqnarray}
Direct evaluation of Eq.~(\ref{Eq:CalphaLast}) gives $\mathcal{C}_\pm = \pm 1$. The total Hall conductivity of the two quasihole bands is hence $\sigma_\textit{xy} = 0$. This corresponds to the Hall conductivity evaluated in the Mott gap. Bulk-edge correspondence implies that edge modes exist in the gap between the two quasihole bands, Fig.~\ref{Fig:FigEXC1}. The quasiparticle bands have $\mathcal{C}_\pm = \mp 1$, which follows from an analogous calculation. 

Small finite inverse lifetime $\delta$ results in a shift of quasiparticle poles away from the real axis. The results of this subsection for $\mathcal{C}_\alpha$ remain valid as long as a gap exists between quasiparticle bands. Figure~\ref{Fig:FigEXC1} shows that whenever the inverse lifetime $\delta$ is on the order of the kinetic energy strength $t_1$, the Mott gap and the two gaps between excited bands are smeared off. It is therefore necessary to require $\delta \ll t_1$. Moreover, this allows to resolve intra-gap edge modes from bulk states in the density of states.

 The edge modes discussed in this section should be visible either in cold--atom experiments using Bragg spectroscopy \cite{Bissbort, Goldman3} and photoemission spectroscopy \cite{Stewart} or in photonic systems, where  the frequency of the incoming-wave can be adjusted.  As mentioned in Sec.~\ref{Sec:NImodel}, artificial gauge fields have already been synthesized in photonic systems \cite{Hafezi}. The more challenging requirements are photon--photon interactions. However, on a single--cavity level it has been shown that photon--photon interactions can be induced by coupling an off--resonant superconducting qubit to a cavity \cite{Hoffman, Carusotto}. 

\section{Conclusions}
\label{Sec:Conclusions}
 
In this paper we have investigated the bosonic Haldane--Hubbard model at unit filling. By combining several numerical and analytical approaches, we have mapped out the phase diagram as a function of two hopping amplitudes and local interaction and found that it consists of two competing types of superfluid and a Mott insulator supporting local plaquette currents. In particular, we found using methods beyond mean-field theory that there is a reentrant transition into the Mott insulator.  We have discussed two distinct superfluid ground states. These are connected either by a first order transition in the weakly interacting regime, or via two second order Mott insulator transitions in the strongly interacting regime.  Different physical properties of the phases are reflected in the ground state density fluctuations and plaquette currents between next--nearest neighbors. All these observables are accessible in present--day ultracold atom experiments. In addition to the study of the ground states, we have 
addressed the excitation spectra in the weakly interacting superfluid and in the Mott domain and found that the corresponding quasiparticle or quasihole excitations consist of bands with non-zero Chern numbers which predict the existence of edge states in the gaps between excited bands.

We expect that our findings can be probed in the near future in ongoing experiments. This work paves the way to open questions about emergent phases at different filling fractions or in multi-component systems, which we will address in future work. For example, related recent studies \cite{Hur20091452, PhysRevB.87.094521, Sheng, Nandkishore, Schaffer1, Schaffer2} of fermions on the half--filled honeycomb lattice have identified emergence of $d$-wave superconducting state close to the Mott transition. Recent Ref.~\cite{CSSF} discusses a chiral spin superfluid phase of two-component bosons in a double-well potential realized on the honeycomb lattice. This shares some features with our proposal but is nevertheless different.

\section{Acknowledgments}
Support by the German Science Foundation DFG
via Sonderforschungsbereich SFB/TR 49, Forschergruppe FOR 801 and the high-performance computing center 
LOEWE-CSC is gratefully acknowledged. AP acknowledges support from the High Performance Computing facilities of the Faculty of Arts and Sciences at Yale University. This work has also been supported from the Labex Palm, Paris-Saclay. We acknowledge discussions with S.~M. Girvin, Markus Mueller, Arun Paramekanti, S.~A. Parameswaran, N.~Regnault, E.~Demler, B.~Halperin, F.~Sols and L. Tarruell. KLH also acknowledges discussions at CIFAR meetings in Canada.
\appendix*

\begin{widetext}
\section{Excitations of the chiral superfluid}
\label{Sec:Appendix1}

In this Appendix, we complement derivations of Sec.~\ref{Subsec:EXCSF}.
In the chiral superfluid at finite values of $t_1$, we have coupling of three momenta ${\bf k}$, ${\bf k}- {\bf K}_A $ and ${\bf k}+ {\bf K}_A$ in Eq.~(\ref{eq:eqset}).
Accordingly, we introduce
\begin{equation}
 \delta\psi({\bf k})=\left(u_{\bf k}^A, u_{{\bf k}- {\bf K}_A}^A, u_{{\bf k}+ {\bf K}_A}^A,  v_{\bf k}^A, v_{{\bf k}- {\bf K}_A}^A, v_{{\bf k}+{\bf K}_A}^A, u_{\bf k}^B, u_{{\bf k}- {\bf K}_A}^B, u_{{\bf k}+ {\bf K}_A}^B,  v_{\bf k}^B, v_{{\bf k}- {\bf K}_A}^B, v_{{\bf k}+ {\bf K}_A}^B\right).
 \label{eq:delpsi}
\end{equation}
 The dispersion relation $\omega({\bf k})$ is obtained by solving the eigenproblem of the $12\times 12$ matrix
\begin{equation}
 h({\bf k}) = \begin{pmatrix} h_{11} & h_{12} \\ h_{21} & h_{22} \end{pmatrix},
 \label{eq:hamcsf}
\end{equation}
where
\begin{equation}
 h_{11}=\begin{pmatrix}
  \delta - d_3({\bf k}) & 0 & 0 & 0  & 0 & n U e^{2 i \phi_A}\\
  0 & \delta - d_3({\bf k}- {\bf K}_A) & 0 & n U e^{2 i \phi_A}  & 0  & 0\\
  0 & 0 & \delta - d_3({\bf k}+ {\bf K}_A)   & 0& n U e^{2 i \phi_A} & 0\\
  0 & -n U e^{-i 2 \phi_A} & 0   &  - \delta - d_3({\bf k})  & 0 & 0\\
    0 &  0&-n U e^{-i 2 \phi_A} & 0 & - \delta - d_3({\bf k}- {\bf K}_A)  & 0 \\
 -n U e^{-i 2 \phi_A} & 0 &  0 & 0 & 0& -\delta - d_3({\bf k}+ {\bf K}_A) 
 \end{pmatrix},
\end{equation}
\begin{equation}
 h_{22}=\begin{pmatrix}
  \delta + d_3({\bf k}) & 0 & 0 & 0  & n U e^{2 i \phi_B}& 0\\
  0 & \delta + d_3({\bf k}- {\bf K}_A) &  0 & 0 & 0 & n U e^{2 i \phi_B} \\
  0 & 0 & \delta + d_3({\bf k}+ {\bf K}_A)& n U e^{2 i \phi_B} & 0  & 0\\
  0 &  0&  -n U e^{-i 2 \phi_B}  &  - \delta + d_3({\bf k})  & 0 & 0\\
  -n U e^{-i 2 \phi_B} &  0 & 0 & 0&  - \delta + d_3({\bf k}- {\bf K}_A)  & 0 \\
 0& -n U e^{-i 2 \phi_B} &  0  & 0 & 0& -\delta +d_3({\bf k}+ {\bf K}_A)
 \end{pmatrix},
\end{equation}
\begin{eqnarray}
 h_{12} &=& \text{diag}(-d_1({\bf k})+ i d_2({\bf k}), -d_1({\bf k}- {\bf K}_A)+ i d_2({\bf k}- {\bf K}_A), -d_1({\bf k}+ {\bf K}_A)+ i d_2({\bf k}+ {\bf K}_A),\nonumber\\ & & d_1({\bf k})- i d_2({\bf k}), d_1({\bf k}- {\bf K}_A)- i d_2({\bf k}- {\bf K}_A), d_1({\bf k}+ {\bf K}_A)- i d_2({\bf k}+ {\bf K}_A)) ,
\end{eqnarray}
and
\begin{eqnarray}
h_{21} &=& \text{diag}(-d_1({\bf k})- i d_2({\bf k}), -d_1({\bf k}- {\bf K}_A)- i d_2({\bf k}- {\bf K}_A), -d_1({\bf k}+ {\bf K}_A)- i d_2({\bf k}+ {\bf K}_A),\nonumber\\ & & d_1({\bf k})+i d_2({\bf k}), d_1({\bf k}- {\bf K}_A)+ i d_2({\bf k}- {\bf K}_A), d_1({\bf k}+ {\bf K}_A)+ i d_2({\bf k}+ {\bf K}_A)).
\end{eqnarray}
Here $\delta = 2n U-\mu =  U n + 3\sqrt{3} t_2$. 
\end{widetext}
\begin{widetext}
The characteristic polynomial of the matrix (\ref{eq:hamcsf}) is too long to be written down completely, so we explicitly show only few most interesting terms:
\begin{eqnarray}
 p(x)&=&x^{12} + x^{10} \left(6 n^2 U^2-6 \delta^2-2\left( {\bf d}({\bf k})^2+ {\bf d}({\bf k}-{\bf K}_A)^2+ {\bf d}({\bf k}+{\bf K}_A)^2\right)\right) + \ldots +() x^2+\delta^{12}\nonumber \\
 &-&2  n^6 U^6 \cos (6 \phi) \left(d_1({\bf k})^2+d_2({\bf k})^2\right)\left(d_1({\bf k}-{\bf K}_A)^2+d_2({\bf k}-{\bf K}_A)^2\right)\left(d_1({\bf k}+{\bf K}_A)^2+d_2({\bf k}+ {\bf K}_A)^2\right).
\label{eq:polynom}
 \end{eqnarray}
\end{widetext}

%

\end{document}